\documentclass{aa} 

\usepackage{graphicx}
\usepackage{txfonts}
\usepackage{amsmath}
\usepackage[colorlinks=true, citecolor = blue]{hyperref}
\usepackage{subfigure}
\usepackage[version=4]{mhchem}

\begin{document} 

   \title{Dissociation and destruction of PAHs and PAH clusters induced by absorption of X-rays in protoplanetary discs around T\,Tauri stars}
   \titlerunning{X-ray induced dissociation and desorption of PAHs and PAH clusters} 
   \subtitle{}
   \author{K. Lange
          \inst{1},
          C. Dominik
          \inst{1}
          \and
          A. G. G. M. Tielens
          \inst{2,}\inst{3}}

   \institute{Anton Pannekoek Institute for Astronomy, University of Amsterdam,
              Science-Park 904, 1098 XH Amsterdam, Netherlands;
              \email{c.dominik@uva.nl}
         \and
             {Leiden Observatory, Leiden University, P.O. Box 9513, 2300 RA Leiden, Netherlands}
        \and {Astronomy Department, University of Maryland, MD 20742, USA}
             }

   \date{Received xxx; accepted xxx}

  \abstract {Only 8\,\% of the protoplanetary discs orbiting a
    T\,Tauri star show emission features of polycyclic aromatic
    hydrocarbons (PAHs). Their abundance is therefore little known. As
    PAHs are strong absorbers of UV radiation, they contribute to the
    heating of the disc photosphere, shielding of UV radiation that
    drives photo-chemistry in the disc, and their abundance is a key
    parameter for determining the strength of photo-evaporative disc
    winds. Soon, high-quality data obtained with the James Webb Space
    Telescope (JWST) will become available with new data to
    interpret.}
    {We want to understand the photochemical evolution of
    PAHs in protoplanetary discs around T\,Tauri stars, and thus
    explain the absence of PAH features. We want to determine whether
    PAHs are destroyed because of the X-ray emission from their host
    stars or whether PAHs can withstand these conditions.}
    {We
    developed a model for the absorption of X-rays by PAHs. X-rays
    with more energy than the K edge of carbon doubly ionise PAHs and
    vibrationally excite them by $\approx15-35$\,eV. With a Monte
    Carlo model, we modelled the dissociation of H, \ce{H2}, and
    \ce{C2H2} from PAH monomers. Furthermore, we modelled the
    dissociation of PAH clusters and the desorption of PAH clusters
    from dust grains caused by X-ray excitation.}
    {We find that small
  PAH clusters quickly desorb and dissociate into individual
  molecules. PAH molecules experience rapid loss of H and acetylene
  \ce{C2H2} by the high excitation and lose \ce{C2H2} on average after
  three X-ray excitations. However, large PAH clusters (coronene
  \ce{C24H12}: 50 cluster members, circumcoronene \ce{C54H18}: 3
  cluster members) can stay intact and frozen out on dust grains.}
    {Based on our results, we expect a gas-phase PAH abundance that is
      lower than 0.01 times the ISM abundance and that rapidly
      decreases over time due to the dissociation of small clusters
      that are subsequently destroyed.  To maintain a higher
      abundance, replenishment processes such as vertical mixing must
      exist. Large PAH clusters remain in the disc, frozen out on dust
      grains, but barely emit PAH features because of their strong
      thermal coupling to dust grains.}


   \maketitle

\section{Introduction}
Almost 50 years ago, unidentified aromatic infrared bands (UIBs) were
seen for the first time in the planetary nebula NGC 7027
\citep{Russell1977} and were then attributed to carbon-rich aromatic
molecules commonly referred to as polycyclic aromatic hydrocarbons
(PAHs) \citep{Duley1981, Leger1984, Sellgren1984, Allamandola1985}.
The typical PAH infrared signals we observe occur at wavelengths of
3.3\,$\mu$m, 6.2\,$\mu$m, 7.7\,$\mu$m, 8.6\,$\mu$m, and 11.3\,$\mu$m.
Other weaker PAH features exist, but are much harder to detect.  Each
of the spectral features can be attributed to a specific vibrational
mode of the chemical bonds: stretching and bending modes between
carbon-carbon bonds and carbon-hydrogen bonds\citep{Tielens2008}.  To
date, PAHs have been identified in numerous astrophysical environments
such as galaxies, the interstellar medium, nebulae, protoplanetary
discs, and the Solar System \citep{Leger1986, Desert1988,
  Verstraete1996, Peeters2002, Acke2004, Tielens2008, Li2009,
  Kamp2011, Tielens2013, Li2020}.  In the context of planet-forming
discs, several surveys have been executed to investigate the infrared
spectra of protoplanetary discs.  Using the infrared space observatory
(ISO), \citet{Acke2004} investigated the spectra of 46 Herbig Ae/Be
stars and found PAH signals in 57\,\% of the discs.  In 2003, the
Spitzer Space Telescope was launched with the InfraRed Spectrograph
(IRS) on board.  \citet{Geers2006} obtained spectra for 38 T\,Tauri
discs and determined a lower limit of 8\% for the detection rate of
PAHs. A few years later, \citet{Acke2010} analysed 53 Herbig Ae
stars with Spitzer and detected PAH features in 70\,\% of them.
Finally, \citet{Valegard2021} analysed Spitzer spectra for
intermediate-mass T\,Tauri stars and detected PAHs in 40\,\% of them.
This difference in detection rate between low-mass and
intermediate-mass pre-main-sequence (PMS) stars clearly shows that the
stellar type affects the evolution and observability of PAHs.
\citet{Seok2017} re-investigated data of 14 T\,Tauri and 55 Herbig
discs and fitted the continuum-subtracted PAH spectra with emission
models from \citet{Draine2007}.  The authors found a positive
correlation between PAH size and stellar effective temperature that
can be explained by the destruction of small PAHs in discs with strong
stellar UV radiation.  Furthermore, the authors did not find a
correlation between disc age and PAH size and conclude that a
replenishment mechanism must act in protoplanetary discs.\\ \\In the
literature, attempts have been made to explain the lower abundance of
PAHs in protoplanetary discs.  \citet{Geers2009} discussed several
different explanations based on their observational findings.  On the
one hand, they considered non-destructive mechanisms such as
coagulation into clusters or the freezing out of PAHs as a realistic
explanation for the lower abundance of observed PAHs.  In these
processes, PAHs are not destroyed, but are harder to detect through
broader emission features by clusters \citep{Allamandola1989} or
thermal coupling to the dust grains.  Both processes were addressed by
\citet{Lange2023} and connected to the strength of the vertical
turbulence in the disc.  On the other hand, destruction mechanisms
such as organic chemistry or ice chemistry may modify frozen-out or
trapped PAHs \citep{Bouwman2011b, Bouwman2011} and deplete them in the
disc.\\ \\Discs around T\,Tauri stars seem to lack PAH emission
features compared to the interstellar medium (ISM) and Herbig star
discs.  Herbig Ae/Be stars are pre-main-sequence stars with masses
between $1.5\,M_\odot \leq M_* \leq 10\,M_\odot$ and have a spectral
type of A or B; thus, they show significant thermal emission in the
ultraviolet (UV) (see \citet{Waters1998} and \citet{Brittain2023} for
reviews on Herbig stars).  In contrast, T\,Tauri stars are low-mass
PMS stars with masses $M_* \leq 3\,M_\odot$ and spectral types of F or
later \citep{Appenzeller1989}.  T\,Tauri stars have much lower stellar
emission at UV wavelengths due to their lower effective stellar
temperature.  Herbig stars have weak X-ray luminosity compared to
their stellar luminosity ($ 10^{-6} \leq L_\text{X}/L_* \leq 10^{-4}$,
\citet{Brittain2023}).  Consequently, UV photons are the main energy
sources for PAHs.  On the other hand, T\,Tauri stars are sources of
strong X-ray emission ($10^{-5}\leq L_\text{X}/L_* \leq 10^{-2}$,
\citet{Preibisch2005b}), presumably caused by their strong magnetic
fields.  T\,Tauri stars are not fully convective, and their
differential rotation speed can explain the strong correlation between
their rotation period and X-ray luminosity \citet{Pallavicini1981} for
the creation of the stellar magnetic field.\\ \\We propose that the
rare (compared to absorbed UV photons in Herbig discs) but very
energetic excitations through absorbed X-rays have a major impact on
the evolution of PAHs in T\,Tauri discs.  The interaction of
high-energy photons, for example far-ultraviolet (FUV), extreme
ultraviolet (EUV), and X-rays, with PAHs was investigated by
\citet{Siebenmorgen2010} and \citet{Siebenmorgen2012}. In these works, the lifetime of PAHs was estimated based on the
amount of time PAH molecules can spend in the photosphere of the
protoplanetary disc.  The authors related the survivability and
observability of the PAHs to the strength of turbulent motions and the
presence of a halo around the disc that can obscure the PAH features.
For this work we re-investigated the interaction of PAHs and X-rays
under consideration of PAH microphysics, which limits the available
excitation energy to a few tens of electronvolts and causes PAHs to become doubly ionised.  In addition, we
also investigate the interaction of gas-phase PAH clusters and
adsorbed PAH clusters with X-rays.  We propose mechanisms to explain
observations of reduced PAH abundances and detection rates of PAH
signals in T\,Tauri discs by the fast destruction of PAH molecules and
small PAH clusters, and the survival of large PAH clusters against
hard radiation to suppress the observable PAH features.  We examine
the ability of X-rays to detach PAHs from dust grain surfaces, to
dissociate PAH clusters into PAH molecules, and to destroy individual
PAH molecules by triggering the ejection loss of acetylene \ce{C2H2}
molecules.\\ \\This manuscript is structured as follows. In
Sect. \ref{sec:methods} we describe our model for the excitation of
PAHs by X-rays and the resulting dissociation cascade of PAH molecules
and PAH clusters.  Then in Sect. \ref{sec:results} we outline how we
applied the model to gas-phase PAHs, gas-phase PAH clusters, and
adsorbed clusters to determine the rate at which the carbon skeleton
of the PAHs is irreversibly destroyed.  Thereafter, we discuss our
findings in Sect. \ref{sec:discussion} in the context of previous
theoretical and laboratory work on the dissociation through X-rays as
well as the implications of our findings for the survival of PAHs and
their state (adsorbed, gas-phase) in T\,Tauri discs.  Finally, we
conclude and present our results in Sect. \ref{sec:conclusions}.

\section{Methods}
\label{sec:methods}
In previous works, the excitation and relaxation of PAHs under irradiation of UV photons was investigated by authors such as \citet{Bakes2001} or \citet{Visser2007}.
After absorption of a UV photon, the entire energy of the photon can be transferred into vibrational excitation that can trigger dissociation of the PAH.
Alternatively, the PAH can ionise and eject an electron.
In this work we consider that X-rays do not deposit their entire energy into vibrational modes, and thus require a more complex treatment.
Therefore, we explain in this section the excitation mechanism of PAHs by X-rays followed by the different dissociation mechanisms (dissociation of the PAH and PAH clusters) that we consider at the given excitation energies.

\begin{figure}
    \centering
    \includegraphics[width=0.99\linewidth]{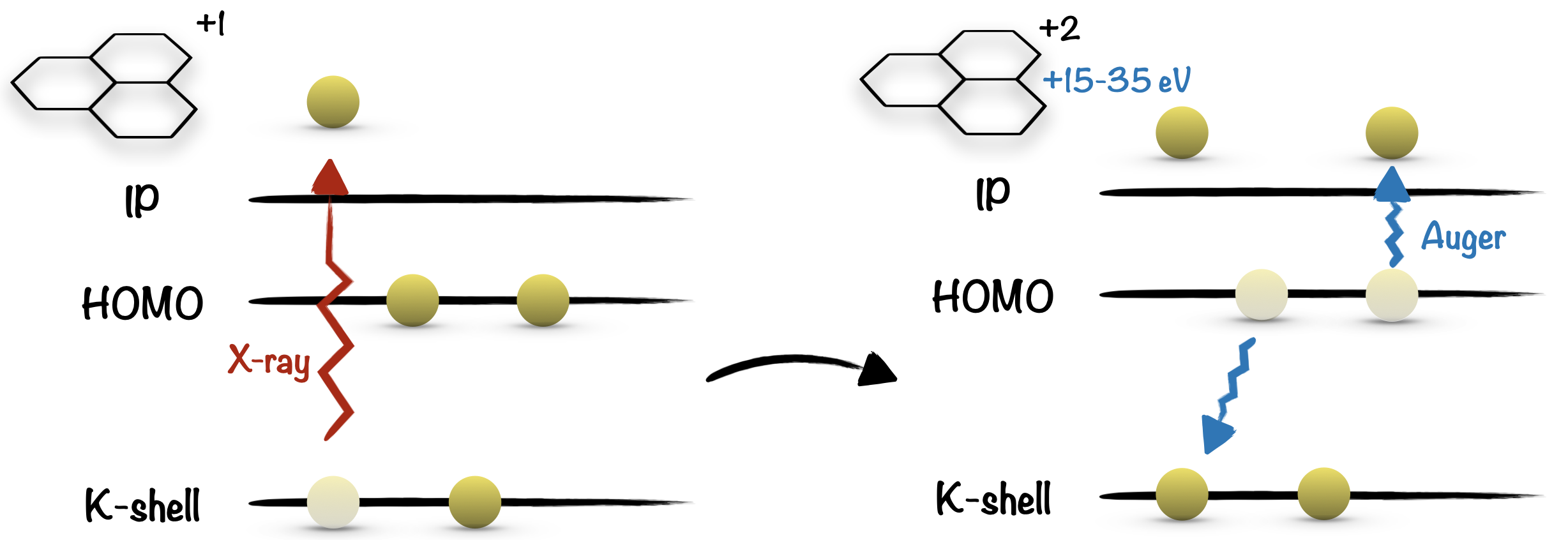}
    \caption{Excitation and ionisation process of a PAH after absorption of an X-ray. The X-ray   ejects an electron from the innermost shell of a carbon atom. Then, an electron from a higher shell fills the vacant inner shell so that the freed energy ejects another electron. A fraction of the energy is carried away by the electron, the remaining fraction is transferred to vibrational excitation of the PAH. In the end, the PAH is  doubly ionised and vibrationally excited by the X-ray absorption.}
    \label{fig:aug_process}
    \vspace{-0.3cm}
\end{figure}

\begin{figure}[h!]
    \centering
    \includegraphics[width=0.8\linewidth]{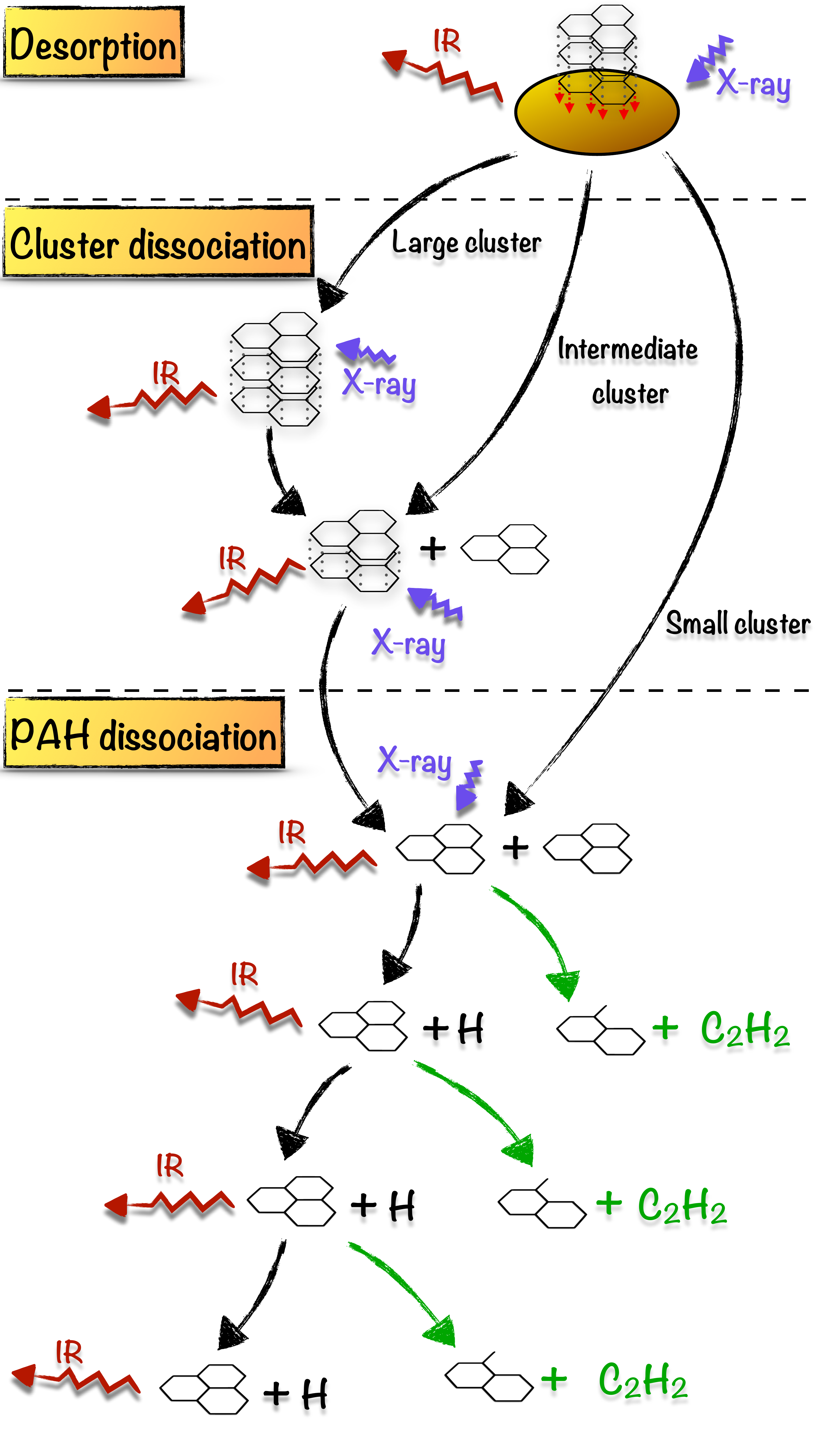}
    \caption{Sketch of cascading processes for PAHs in protoplanetary discs. X-rays can trigger evaporation of PAH clusters from dust grains or the vibrational energy of the cluster is transported to the dust grains and emitted as thermal radiation. Dependent on the cluster size and PAH species, the evaporated cluster can be fully destroyed into monomers (small cluster), can lose a few of its monomers (intermediate clusters), or not lose any monomer (large clusters). Then, the subsequent absorbed X-rays break down the cluster until the entire cluster has been dissociated into molecules. Finally, the absorption of X-rays by individual PAH molecules triggers dissociation of H and \ce{H2,} which cools the PAH without structural damage, followed by IR cooling. Simultaneously, acetylene (\ce{C2H2}) fragments can be ejected from the PAH, damaging the carbon backbone, which is difficult to repair without complex chemistry.}
    \label{fig:cascadesketch}
    \vspace{-0.35cm}
\end{figure}

\subsection{PAH interaction with X-rays}
As with all particles found in astrophysical environments, PAHs are exposed to high-energy photons, including X-rays, especially regarding circumstellar discs surrounding T\,Tauri stars. 
The interaction between PAHs and high-energy photons is crucial as X-ray photons can deposit a large amount of energy into individual molecules and molecular clusters.
X-rays interact with the K-shell electrons of the carbon atoms and the resulting (free) electron carries some of the excess energy away as kinetic energy (Fig. \ref{fig:aug_process}, left).
Subsequently, a valence electron drops down to the vacant inner K shell.
The energy associated with this electron transition may be released as an X-ray photon or leads to the ejection of a second electron, the Auger electron \citep{Auger1923}, causing double-ionisation of the PAH (Fig. \ref{fig:aug_process}, right).
Generally, Auger-Meitner ionisation dominates over X-ray fluorescence for low atomic numbers $Z$, such as for carbon \citep{Bambynek1972}. 
A fraction of the electronic excitation energy is carried away by the Auger electron, but the remaining energy difference is transferred into vibrational excitation of the molecule.
The amount of available excitation energy depends on which orbital the electron that fills the K shell comes from.
Hence, a distribution of possible excitation energies is expected.
We assume that 15-35 eV is left behind as internal excitation energy \citep{Micelotta2010}, which agrees with the conclusions of \citet{Reitsma2015} that the vibrational excitation follows a distribution around 20\,eV based on the Auger spectrum of benzene, whichg  has been well studied in experiments and theory \citep{Rye1984, Jayadev2023}. To approximate the absorption cross-section of different PAH species, we adopted the elemental carbon cross-section from \citet{Biggs1988} approximated by fourth-order hyperbolic functions where the coefficients for different energy intervals are given in Table \ref{tab:C_coefficients}:
\begin{equation}
    \sigma_i = \frac{A_{1,i}}{E} + \frac{A_{2,i}}{E^2} + \frac{A_{3,i}}{E^3} + \frac{A_{4,i}}{E^4} \text{\,.}
    \label{eq:sigma_C}
\end{equation}
\begin{table*}[]
    \centering
    \caption{Coefficients for Equation \eqref{eq:sigma_C} from \citet{Biggs1988} to calculate the elemental carbon absorption cross-sections for high-energy photons.}
    \begin{tabular}{lrrrr}
    \hline
         Interval $i$ & $A_{1}$ & $A_{2}$ & $A_{3}$ & $A_{4}$  \\ \hline\hline
         0.01\,keV $\leq E \leq $ 0.0457\,keV & $5.704 \cdot 10^3$ & 0 & 0 & 0 \\
         0.0457\,keV $\leq E \leq $ 0.284\,keV & $-3.935 \cdot 10^2$ & $-3.219 \cdot 10^2$ & $-6.549$ &  $2.086 \cdot 10^{-1}$\\
         0.284\,keV $\leq E \leq  $ 0.8\,keV & $-9.022 \cdot 10^2$ & $1.76 \cdot 10^3$ & $1.549 \cdot 10^3$ & $-2.28 \cdot 10^2$\\
         0.8\,keV $\leq E \leq $ 4\,keV & $-7.363$ & $-1.537 \cdot 10^1$ & $2.672 \cdot 10^3$ & $-4.482 \cdot 10^2$ \\
         \hline
    \end{tabular}
    \label{tab:C_coefficients}
\end{table*}
\\The commonly used X-ray cross-sections of \citet{Verner1995} produce similar results. 
We scale the carbon cross-section with the number of carbon atoms in each PAH to approximate the PAH absorption cross-section in the X-ray photon regime.
The findings agree with the absorption cross-section near the K edge measured by \citet{Boechat-Roberty2009} after scaling the absorption cross-section to the number of carbon atoms.
We note that through this approximation we neglected features around the K edge arising from the molecular structure of PAHs, as investigated in the laboratory by  \citet{Reitsma2015} or \citet{Huo2022}, among others.
However, this is only a small interval compared to the whole considered X-ray spectrum between 280\,eV and 2\,keV that we consider in our study following \citep{Siebenmorgen2010}. 
Therefore, the molecular fine-structure of the X-ray absorption spectrum around the K-edge is negligible in our calculations as the deposited vibrational excitation energy by the X-ray does not depend on the X-ray energy itself. 
In principle, the cascade model can also consider higher-energy X-rays, as the energy deposited into the PAH is independent from the energy of the X-ray photon. However, implementing such a model would require a more complex vertical structure to determine the penetration depth of all X-rays at varying disc heights.\\
\\To estimate the number of emitted X-rays by the star, we followed the model of \citet{Siebenmorgen2010} and assumed an X-ray luminosity of $L_\text{X}/L_* \approx 2.5 \cdot 10^{-4}$ based on studies by \citet{Preibisch2005} who used observations of the Chandra X-ray Observatory of T\,Tauri stars in the Orion Nebula Cluster.
We take the T\,Tauri star BP\,Tau \citep{Grankin2016} as a typical representative of the T\,Tauri stars.
BP Tau has a stellar luminosity of $0.88$\,$L_\odot$ and a stellar effective temperature of 4040\,K.
As a result, our model assumes\footnote{We would like to clarify that this does not take into account potential measured X-ray luminosities from BP\,Tau specifically.} an X-ray luminosity of $L_\text{X}=8.4\cdot10^{29}$\,erg/s for a BP\,Tau-like T\,Tauri star for X-rays between 280\,eV and 2\,keV.
We note that the X-ray luminosity of T\,Tauri stars can be in the range  $10^{-5}\leq L_\text{X}/L_* \leq 10^{-2}$, \citet{Preibisch2005b}, so all of our calculated rates and  timescales must be adjusted and scaled appropriately.
We assume that the X-ray photon flux is emitted from a hot black body, so the X-ray frequency range lies in the Rayleigh-Jeans limit of the black body and can be approximated by $F(\nu) \propto \nu^2$.
The absorption rate of X-rays can then be calculated by integration of the product of the absorption cross-section $\sigma(E)$ and X-ray photon flux $F_x(E)$:
\begin{equation}
    k_\text{X} = \int \sigma(E)F_x(E) dE.
\end{equation}

\subsection{Dissociation cascade model}
The absorption of an X-ray leaves a PAH highly excited. 
Depending on its state (cluster, gas-phase, frozen-out), the PAHs have multiple channels to cool down after excitation.
An overview of the different evolutionary pathways and their competing cooling processes is displayed in Fig. \ref{fig:cascadesketch}.
We consider non-dissociative cooling through emission of IR photons with a typical rate of $r_\text{cool} = 1$\,s$^{-1}$ \citep{Andrews2016} if the PAH is in the gas phase.
If the PAH or PAH cluster is frozen on a grain, the estimated heat transfer to the dust grain ($r_\text{cool} = 10^{12}$\,s$^{-1}$) \citep{Lange2023} is faster than the emission of IR photons by the PAH.
We model the heat transfer to the dust grain by assuming that vibrational coupling between the PAH and the dust grain surface will be similar to graphite conduction through the c plane between the individual layers described by
\begin{equation}
    \frac{\text{d}Q}{\text{d}t} = \frac{\kappa A \Delta T}{\Delta d}
    \label{eq:conduction}
,\end{equation}
where the contact surface $A$ is determined by the area of a flat PAH molecule that is attached to the grain, and
\begin{equation}
    A = \pi \left(\sqrt{N_\text{C,0}} \cdot 0.9 \text{\AA}\right)^2 = N_\text{C,0} \cdot 2.5 \cdot 10^{-16} \text{\,cm}^2
    \label{eq:surfaceareaPAH}
,\end{equation}
where $N_\text{C,0}$ is the number of C atoms of a PAH monomer \citep{Tielens2008}, $\kappa = 0.04$\,W/m/K = 4\,000 erg/s/cm/K is the heat conduction coefficient of graphite \citep{Alofi2014}, and $\Delta d$ is the contact distance between two graphite layers that we assume to be $\Delta d = 3.34\,\AA$ \citep{Andres2008}.\\
\\For dissociative cooling, we consider the release of H atoms, the release of acetylene (\ce{C2H2}), and the ejection of entire PAH molecules if clusters are considered.
The fragmentation of a few small PAHs has been studied experimentally by Ling, Lifshitz, and co-workers on timescales of several microseconds \citep{Ling1995,Ling1998}.
At these timescales, the loss of H, \ce{H2}, and \ce{C2H2} has been fitted by simple Arrhenius laws and the parameters for the rates involved have been derived. However, it is unclear whether these rates can be extended to the much longer timescales associated with radiative cooling ($\approx0.1$\,s), the much higher internal energies involved in X-ray irradiation, and the much larger PAHs considered here.
In particular, we recognise that laboratory experiments imply that large PAHs (more than 32 carbon atoms) are first stripped of all their Hs before the resulting (isomerised) carbon clusters shrink through \ce{C2} loss \citep{Zhen2014}.
In the absence of direct information on the specific microphysical steps in the chemical reaction routes that would allow such an extrapolation, we adopted the Arrhenius law derived from the time-dependent fragmentation studies on pyrene (\ce{C16H10}).
We expect that this choice overestimates the breakdown of the carbon skeleton through loss of \ce{C2H2}. 
Then, dissociative processes can be approximated by an Arrhenius expression of the   form
\begin{equation}
    k = k_0 \cdot \text{exp}\left(\frac{E_\text{A}}{-k_\text{B}T_\text{e}}\right)
    \label{eq:arrhenius}
\end{equation}
to estimate the dissociation rate given an activation energy $E_\text{A}$.
The pre-exponential factor can be calculated through
\begin{equation}
    k_0 = \frac{k_\text{B}T_\text{e}}{h}\text{exp}\left(1+\frac{\Delta S}{R}\right)
,\end{equation}
where $k_\text{B}$ is the Boltzmann constant, $T_\text{e}$ the effective temperature of the PAH, $h$ the Planck constant, $R$ the ideal gas constant, and $\Delta S$ the change in entropy of the reaction \citep{Tielens2021}.
We used values available in the literature for pyrene (\ce{C16H10}) even though we note that the exact value depends on the PAH molecule considered.
For \ce{H2} loss we assumed $E = 3.52$\,eV and $\Delta S = -12.7$\,cal/K/mole \citep{Ling1995}, $E = 4.6$\,eV and $\Delta S = 10.7$\,cal/K/mole \citep{Ling1995} for H loss, and $E = 4.4$\,eV and $\Delta S = 3.5$\,cal/K/mole \citep{Ling1998} for \ce{C2H2} loss.
In all dissociative cases, the PAH loses the binding energy $E' = E - E_\text{A}$ and additionally a fraction of the energy that is carried away by the ejected atom from the PAH or ejected molecule from the cluster.
In the case of atom loss, the additional energy that is carried away by the fragments as kinetic energy is of the order of $2\,k_\text{B}T,$ which is much smaller than the total energy of the system.
In the case of PAH clusters, a fraction of the vibrational energy is carried away by the ejected molecule that is calculated by the fraction of the degrees of freedom that are lost
\begin{equation}
    E_\text{loss} = \frac{3N_\text{mon}-6}{3(N_\text{dau}+N_\text{mon})-6} E'
    \label{eq:energy_fraction}
,\end{equation}
where $N_\text{dau}$ is the number of atoms in the daughter cluster and $N_\text{mon}$ is the number of atoms in the ejected PAH.
We assumed $k_0 = 2.5\cdot10^{17}$\,s$^{-1}$ for all PAH clusters similar to the dissociation of PAH clusters derived in the Appendix of \citet{Lange2021}.\\
\\To calculate the excitation temperature of the PAHs and PAH clusters, we used the energy-temperature relation \citep{Bakes2001, Tielens2021}
\begin{equation}
    T_\text{m} = \begin{cases}
    3750 \left(\frac{E\text{(eV)}}{3N-6}\right)^{0.45}\text{\,K}\text{\hspace{1cm} if $T_m < 1000$\,K}\\
    11000 \left(\frac{E\text{(eV)}}{3N-6}\right)^{0.8}\text{\,K}\text{\hspace{1cm} if $1000\text{\,K} < T_m$}\\
    \end{cases}
    \label{eq:T(E)}
,\end{equation}
for which the heat-bath correction needs to be applied to obtain the necessary effective temperature $T_\text{e}$ in Equation \eqref{eq:arrhenius}:
\begin{equation}
    T_\text{e} = T_\text{m} \left(1-0.2\frac{E_\text{A}}{E}\right) \text{\,.}
    \label{eq:Teff}
\end{equation}
Finally, the probability of each of the dissociative and non-dissociative cooling events can be calculated by evaluating the rate of each dissociation and cooling event over the total rate of events:
\begin{equation}
    p_i(E) = \frac{k_i(E)}{\sum_i k_i(E)}.
    \label{eq:prob}
\end{equation}
Using a random number generator with the given probability for each event, we can consider all paths for cooling with a Monte Carlo scheme and determine the probability for each of the dissociative losses.
We followed the cascade of cooling until the PAH cooled down to the initial temperature or was damaged at the carbon backbone by the release of \ce{C2H2}.
For gas-phase PAH monomers, we followed $10^6$ Monte Carlo steps, which roughly corresponds to  $10^4-10^5$ modelled photon absorptions. 
For gas-phase PAH clusters, we used $10^7$ Monte Carlo steps, which corresponds to $10^5$ photon absorptions.
For an adsorbed PAH cluster, we used $\approx 10^9$ Monte Carlo steps, which  corresponds to $\approx 10^6$ X-ray photon absorptions.
\begin{figure}
    \centering
    \includegraphics[width=1\linewidth]{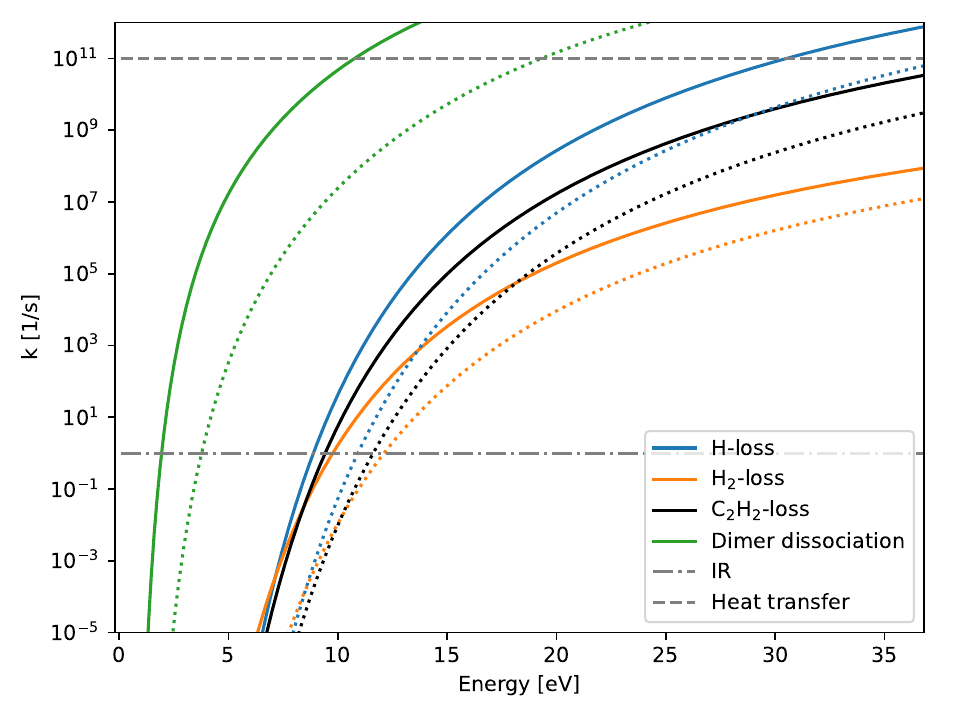}
    \caption{Comparison of dissociation rates for coronene C$_{24}$H$_{12}$ (solid lines) and ovalene C$_{36}$H$_{18}$ (dashed lines). Only for low energies where H and C dissociation starts to appear do H$_2$ and C$_2$H$_2$ loss have comparable rates to H loss;  for high energies H loss always dominates. Dissociation of a dimer is always favoured because of the much weaker binding energy between two PAHs. For ovalene, the reactions appear at nearly the same temperature but at higher energies because of the larger heat capacity. Hence, the increasing size of PAHs and PAH clusters increases their stability against X-rays as the deposited energy  is independent of the structure. The dissociation rate for clusters additionally depends on the size of the cluster. The typical rates for IR cooling and energy transfer to dust grains are indicated by the grey lines.}
    \label{fig:rates}
    \vspace{-0.3cm}
\end{figure}
\section{Results}
\label{sec:results}
Our analysis is structured in the following way: first we apply the cascade model to individual gas-phase PAHs and investigate the loss of acetylene and hydrogen induced by X-rays.
Then we extend the model to gas-phase clusters and analyse their stability and destruction after excitation by X-rays.
Finally, we extend our study to adsorbed PAH clusters and their ability to evaporate from the grain surface.\\
\\The goal of our models is to investigate the interaction of PAHs with X-rays after they have potentially clustered and frozen-out in the shielded midplane of the disc.
One of these mechanisms can be vertical mixing, but there are  other processes that can transport PAHs and grains from an optically thick to an optically thin regime.
The height and location in the disc where this occurs is strongly dependent on the structure of the disc, but also on the relative strength between the UV field and the X-ray luminosity.
Our obtained results are therefore very general and  a sophisticated application of such a model to a specific disc to make precise predictions requires proper treatment of radiative transfer and disc morphology.
The transport from a shielded to an exposed regime is much more complicated than our estimate and there might be regions between the midplane and the photosphere where some of our simulated processes are only partially active.
However, it is beyond the scope of this work to treat this in detail.\\
\\All our calculations are performed under full X-ray exposure on the basis of single X-ray absorptions, as multi-photon events are highly unlikely. 
Given the parameters from the X-ray model, for our typical T\,Tauri disc we derive an X-ray absorption rate at a radial distance from the star $r$ in the photosphere of the disc given the number of carbon atoms in the molecule $N_\text{C}$,
\begin{equation}
    k_\text{X}(r) = 1.8 \cdot 10^{-9}  \left(\frac{r}{10\,\text{au}}\right)^{-2} \frac{N_\text{C}}{24}\,\text{s$^{-1}$},
\end{equation}
which corresponds to an absorption of one single X-ray photon every 17.4\,yr for a coronene at a distance of  10\,au. 
All calculations assume that only the optically thin photosphere of the disc is considered so that X-rays reach PAHs without previous scattering or absorption.

\subsection{\ce{C2H2} loss of gas-phase PAH monomers}

\begin{figure}
    \vspace{-0.3cm}
    \centering
    \includegraphics[width=0.99\linewidth]{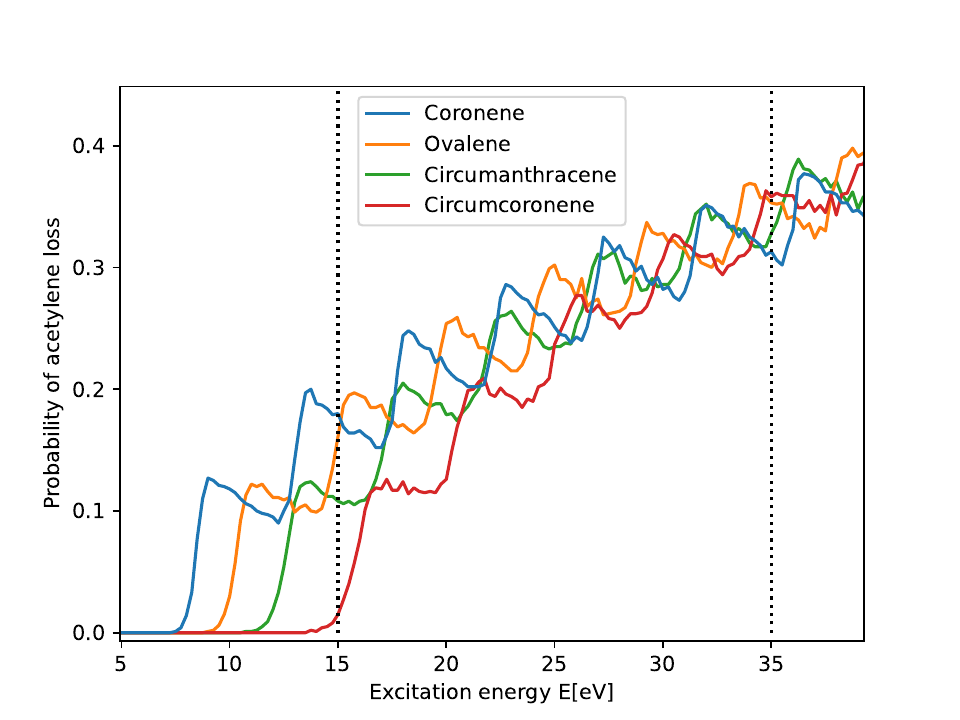}
    \caption{Fraction of destroyed gas-phase PAHs for our four considered PAH sizes (coronene \ce{C24H12} to circumcoronene \ce{C54H18}) after absorption of an X-ray followed by vibrational excitation through the release of an Auger electron.}
    \label{fig:gasphasemonomer}
    \vspace{-0.2cm}
\end{figure}

In our first considered case, we investigated the impact of X-ray absorption on gas-phase PAH monomers.
Our simulations accounted for cooling through infrared photon emission and the dissociation by release of H, H$_2$, and C$_2$H$_2$.
Our results show that the stripping of single hydrogen atoms from the PAH molecule is the most likely event for all energies at which IR cooling is not dominant (see Figure \ref{fig:rates}), which agrees with the appearance energy of 12\,eV measured for coronene (\ce{C24H12}) by \citet{Jochims1994}.
The loss of acetylene only occurs with a probability of a few  per cent at all energies.
Therefore, in most cases, the PAH experiences sequential losses of a few hydrogen atoms until it has cooled down to energies where infrared cooling starts to dominate and the PAH eventually cools down fully.
Then, the PAH has sufficient time to re-hydrogenate by collisions with hydrogen as the collision timescale is of the order of seconds.
If the carbon structure of the PAH is damaged, acetylene loss occurs at all hydrogenation stages, although lower hydrogenation states are slightly preferred (see probability curves of the model in Appendix \ref{appendix:probs}).\\
\\Figure \ref{fig:gasphasemonomer} shows the impact of different amounts vibrational energies left in the molecule after the absorption of X-rays on the dissociation of various PAH species (coronene C$_{24}$H$_{12}$, ovalene C$_{32}$H$_{14}$, circumanthracene C$_{40}$H$_{16}$, and circumcoronene C$_{54}$H$_{18}$).
Our findings show that in the energy range of 15-35\,eV, there is no discernible difference between the average dissociation probabilities of the simulated PAH species.
However, small-scale variations in the loss of acetylene of each PAH species arise from the sequential loss of hydrogen atoms and the subsequent resampling of the model dissociation curves (see Appendix \ref{appendix:probs}).
This sub-structure in the dissociation is the reason why the exact excitation energy (even on a small scale) is important and can change the probability of acetylene loss by a few per cent.
Additionally, we observe a large-scale trend where larger PAHs are slightly more likely to lose acetylene (\ce{C2H2}) because the energy loss  through H loss is independent of the PAH size.
Hence, larger PAHs experience a smaller drop in temperature through H loss, which favours acetylene (\ce{C2H2}) loss slightly more.
Only the lowest energy required for acetylene loss and the energy required to fully dehydrogenate the PAH molecule are affected by the PAH size.
Overall, our study suggests that the probability of damaging the carbon backbone of PAHs by X-rays is only weakly dependent on the considered PAH species in our estimated excitation energy range. Our results suggest that acetylene loss of gas-phase PAHs occurs every few X-ray excitations, so we expect that gas-phase PAHs are (mostly) irreversibly destroyed  of the order of ten times the X-ray absorption timescale.

\subsection{Cluster dissociation of gas-phase PAH clusters}
\label{sec:results_diss}
\begin{figure*}
    \centering \includegraphics[width=1\linewidth]{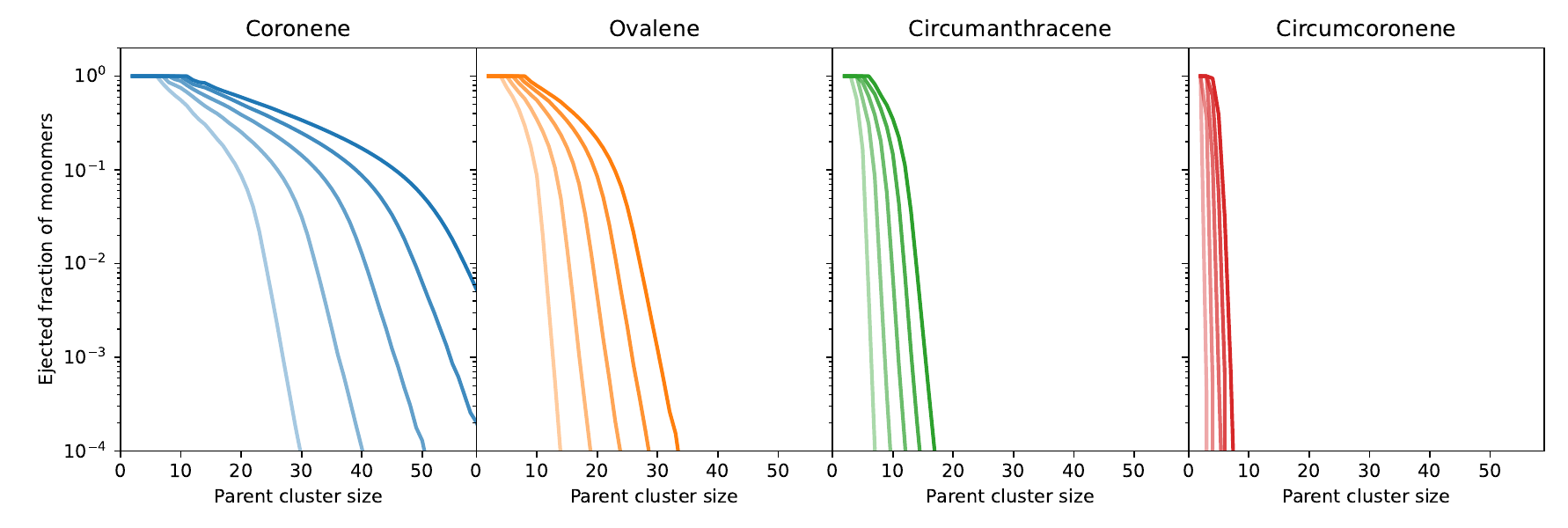}
    \caption{Average fraction of ejected PAH monomers from gas-phase cluster after excitation with 15\,eV, 20\,eV, 25\,eV, 30\,eV, and 35\,eV (from light to dark). With larger cluster size and larger PAH size the probability of losing cluster members strongly decreases so that the cluster is stabilised against X-ray radiation.}
    \label{fig:gasphase_cluster_evaporation}
\end{figure*}

In th enext step we expanded our investigation from gas-phase PAH monomers to PAHs that are present as clusters in the gas phase.
In this scenario, PAH clusters can experience additional ejection of single PAH molecules from the cluster.
This is an additional cooling process that needs to be taken into account as the activation barrier for breaking the van der Waals bond of the cluster is much smaller than the activation barrier for breaking chemical bonds such as acetylene or hydrogen.
For that reason, the evaporation of monomers is much more favoured compared to processes such as the loss of H, H$_2,$ and C$_2$H$_2$.
Hence, the sequential loss of PAH monomers occurs before any chemical bond between hydrogen and carbon atoms is broken.
As a significant amount of energy is carried away by the ejected PAH molecule from the cluster following Equation \eqref{eq:energy_fraction}, the leftover energy of the daughter cluster is hardly ever sufficient to break chemical bonds of the PAHs that have evaporated from the cluster. 
Only for dimers and trimers do the individual monomers retain enough energy to dehydrogenate.\\
\\Figure \ref{fig:gasphase_cluster_evaporation} displays the average number of ejected PAH monomers after a single X-ray excitation as a function of parent cluster size.
As the remaining vibrational excitation energy after the ejection of an Auger electron is unknown, we did the calculation for  energies between 15\,eV and 35\,eV, which is indicated by line colour (from light to dark with increasing excitation energy).
A probability of unity corresponds to the destruction of the cluster into only individual monomers.
We find that for very small clusters with four to six monomers the whole cluster is disrupted by a single X-ray excitation.
For slightly larger clusters, the excitation energy needs to be high enough (30\,eV) to still break the entire cluster down into individual monomers.
Our simulations show that the size of the individual PAH molecules in a cluster is more important for the stabilisation of a cluster than the number of PAH molecules that the cluster consists of. 
For coronene \ce{C24H12}, cluster sizes of around 60 cluster members are needed to fully stabilise the cluster for the typical lifetime of a protoplanetary disc of a few million years.
For circumcoronene \ce{C54H18,}  clusters with six cluster members are stable.\\
\\Our simulations imply that the formation of PAH clusters slows down the process of PAH destruction by fully stabilising large clusters against hard radiation (compared to the lifetime of a protoplanetary disc of $10^6$ years) and slowing down the process of irreversible PAH destruction for small PAH clusters by releasing individual PAH molecules from time to time.

\subsection{Desorption of adsorbed clusters}
\label{sec:results_desorb}
\begin{figure*}
    \centering 
    \vspace{-0.2cm}
    \includegraphics[width=0.99\linewidth]{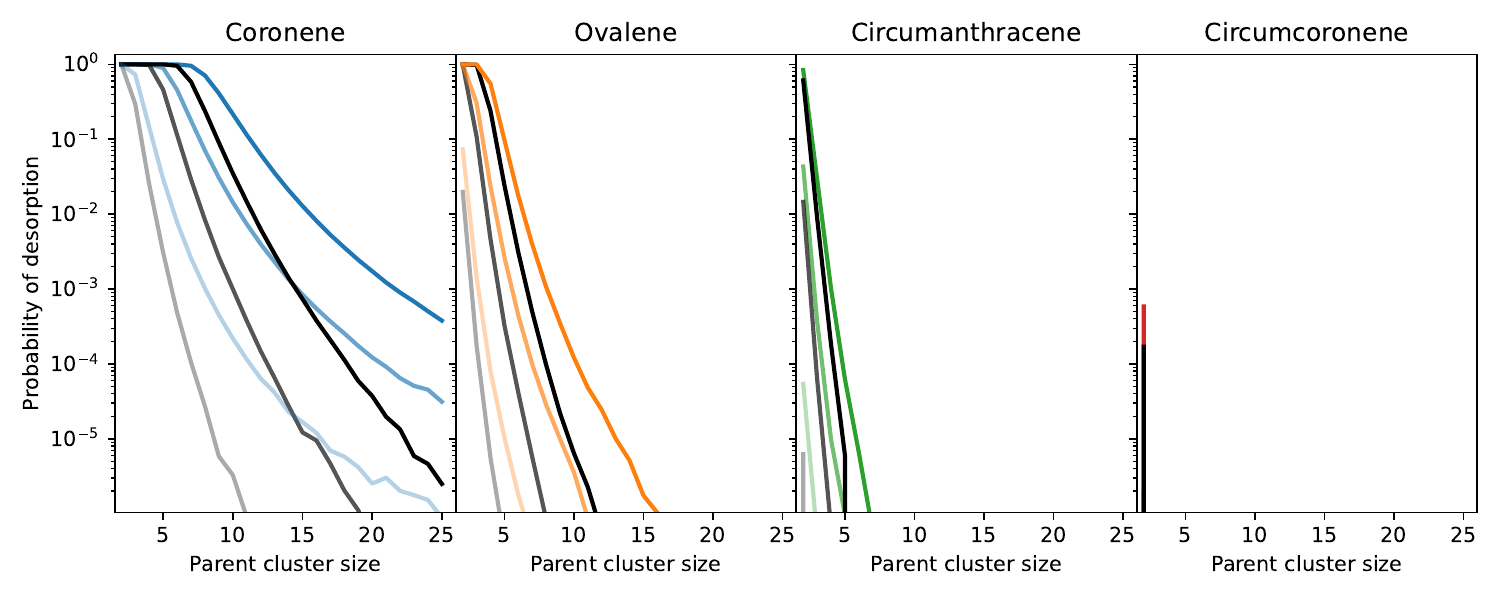}
    \caption{Probability of desorption of a given cluster size from a dust grain at temperature $T=200$\,K (\textit{grey lines}) and $T=400$\,K (\textit{coloured lines}) after absorption of an X-ray for the four considered PAH species between 24 C atoms and 54 C atoms. The different shades of colour  indicate the assumed excitation energy (light: 15\,eV, medium: 25\,eV,  dark: 35\,eV). The considered PAH species has a much larger effect on the desorption probability than the PAH cluster size or dust grain temperature.}
    \label{fig:adsorbed_cluster}
    \vspace{-0.2cm}
\end{figure*}

We next considered the case where PAHs and PAH clusters are adsorbed on dust grains.
Our grains have two different temperatures, warm dust at 400\,K and colder dust at 200\,K.
As the energy transfer to the dust grain is much faster than the IR cooling of gas-phase PAHs, the effective internal energy required for fragmentation processes is raised; hence, the timescale on which these processes must happen is greatly diminished.\\
\\Figure \ref{fig:adsorbed_cluster} shows the probability of desorption for the four PAH species that we consider in our study for two different dust grain temperatures (400\,K: \textit{colours}, 200\,K: \textit{black}).
The increased threshold for evaporation to happen within the cooling timescale heavily impacts the clusters' ability to evaporate from the grain.
We find a   trend similar to that in the dissociation of gas-phase PAH clusters: with increasing size of the PAH species and increasing cluster size, the probability of desorption from the grain decreases drastically.
Only small clusters can immediately evaporate from the grain surface.\\
\\After evaporation from the grain surface, the remaining excitation energy after desorption is high enough to trigger ejection of monomers from the cluster.
The distribution of leftover energies is dominated by the highest possible energy (binding energy of the cluster subtracted from the initial energy) and drops exponentially to lower energies as some clusters cool down before they leave the grain surface.
Hence, depending on the temperature of the dust grain, the desorbed cluster can have leftover energies below and above the initially assumed excitation energy for the gas-phase clusters in the previous section.
Nevertheless, the resulting average fraction of lost monomers after desorption from the grain surface is comparable to that of initial gas-phase clusters excited by X-rays.
Hence, the desorbed clusters experience either partial destruction (larger clusters) or full destruction (small clusters) without absorbing another X-ray photon. We summarise that the desorption of clusters from dust grains is much slower than the destruction of gas-phase clusters and only small clusters with fewer than 25 monomers can desorb from the grain surface in a protoplanetary disc relevant timescale.

\section{Discussion}
\label{sec:discussion}
\subsection{Evolution of PAHs in T\,Tauri discs}
An important question in the search for PAHs in T\,Tauri discs is whether PAHs are absent from most of the T\,Tauri discs or whether they are present in undetectable low abundances compared to Herbig Ae/Be discs, and therefore have only been detected in a few sources.
A survey with the Spitzer space telescope led to inconclusive results, identifying the limiting feature-to-continuum ratio obtainable with Spitzer, and pointing out the necessity of future work \citep{Geers2006}. 
The interaction of X-rays with PAHs covered in this work is only one of the puzzle pieces that is required to understand this observational trend. Disc-substructures in the gas or dust disc have only been partially covered in the literature, and currently  no complete model of PAH chemistry including ionisation, atom-loss, UV- and X-ray chemistry, and (radial and vertical) transport through the disc exists. We urge that  these models be explored in the future to interpret the upcoming JWST spetra of T\,Tauri discs.\\
\\We wanted to combine the results of our calculations using the small PAH coronene \ce{C24H12} as an example to estimate the timescale on which clusters transition from the desorbed state to the irreversible destruction of the carbon backbone of the PAHs to determine the timescale for destruction of PAHs due to X-ray absorption.
Hence, we would like to emphasize our assumed initial disc state.
The protoplanetary disc is in steady state with no remnants of the molecular cloud.
The PAH population that has been inherited from the ISM has had time to form clusters of different sizes. Then PAHs and clusters have frozen-out onto dust grains where it is cold enough.
Now, X-rays are allowed to interact with frozen-out PAHs and frozen-out clusters to observe what effect X-rays have on the gas-phase abundance of PAHs.


\subsubsection*{From frozen out to destroyed}

We wanted  to estimate the amount of adsorbed clusters and gas-phase monomers as a function of time if they are continuously exposed to stellar X-rays.
Therefore, we assumed that coronene \ce{C24H12} is initially present with ISM abundance and then clustered in the protoplanetary disc before being adsorbed on dust grains as described by \citet{Lange2023}.
Integration of the resulting cluster size distribution (Figure 5 in \citet{Lange2023}) shows that 9\,\% of the total PAHs are adsorbed as clusters with 25 cluster members or fewer.
Using the desorption time of the specific cluster sizes, the cluster dissociation time, and the monomer dissociation time, we can estimate the lifetime of the clusters on the dust grains and the resulting fraction of monomers in the gas phase.
The desorption timescale is given by the reciprocal of the desorption probability per X-ray multiplied by the X-ray absorption timescale.
For the dissociation of clusters, we assume that with each X-ray photon absorption a fraction of the monomers enters the gas phase (see Figure \ref{fig:gasphase_cluster_evaporation}).
Step by step, the breakdown of a cluster is calculated until the cluster has completely dissociated into monomers. 
For the destruction of a PAH, we assume that three X-ray absorptions
result in the destruction of a PAH monomer by loss of acetylene
\ce{C2H2} because the average probability of \ce{C2H2} loss is one
third.

\begin{figure}
    \vspace{-0.15cm}
    \centering 
    \includegraphics[width=0.99\linewidth]{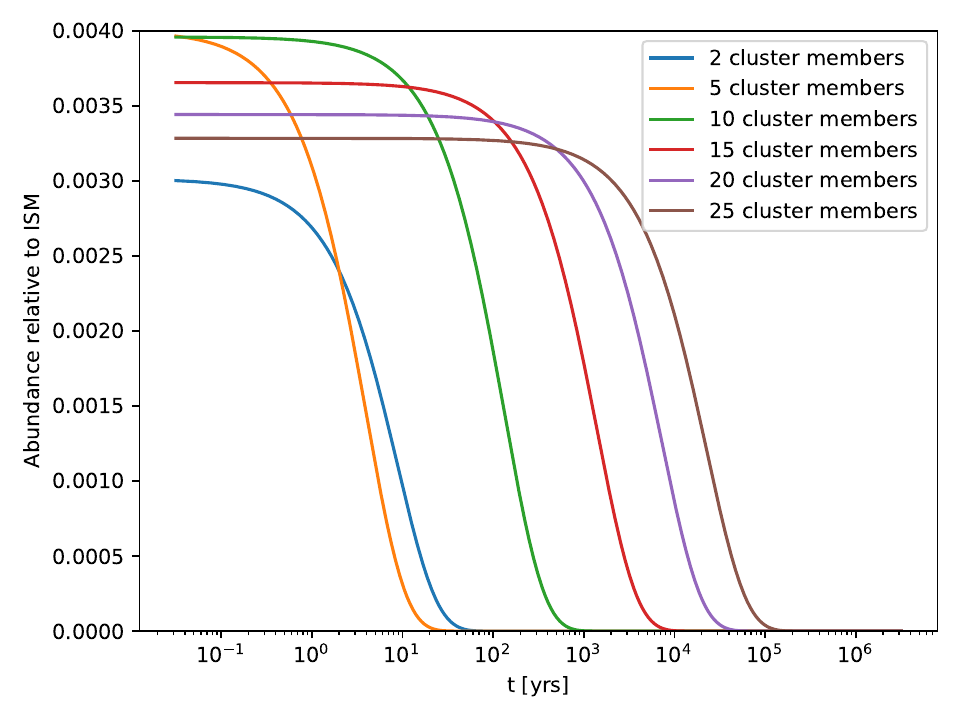}
    \caption{Fraction of initial PAH ISM abundance of each adsorbed cluster size that is desorbed by X-ray photons over time at a 10\,au distance from the central star BP\,Tau (for coronene). A PAH cluster size distribution originating from clustering simulations \citep{Lange2023} and the desorption timescale from Figure \ref{fig:adsorbed_cluster} are assumed.}
    \label{fig:fraction_individually}
    \vspace{-0.25cm}
\end{figure}

Figure \ref{fig:fraction_individually} shows the time evolution of adsorbed coronene clusters smaller than 25 members at 10\,au for the stellar parameters of BP\,Tau.
Small PAH clusters desorb on a single X-ray timescale of a few years, while for clusters larger than a decamer, the timescale for desorption increases to values equivalent to the lifespan of a protoplanetary disc of a few million years.
Thus, the distribution of adsorbed PAH clusters will show fewer and fewer small clusters over time and will be dominated by larger clusters.
Figure \ref{fig:fraction_all} shows the time evolution of adsorbed and gas-phase clusters summed-up over all cluster sizes.
The initial phase of clustering and adsorption is not shown in the figure because the time to cluster and adsorb on dust grains is only a few years at 10\,au.
Initially, small clusters are efficiently desorbed and dissociated into monomers, which are subsequently destroyed by the following X-ray absorption.
Therefore, in the beginning, the PAH abundance reaches its maximum.
Once the reservoir of adsorbed small clusters is depleted and all have desorbed, the slow desorption rate of larger clusters and short destruction timescale of monomers reduces the abundance of monomers, leaving only a small fraction of the original ISM-PAH abundance in the gas phase.
The gas-phase abundance further decreases initially ($t\leq10^3$\,yr) when larger PAH species are considered, even though their increased desorption timescale preserves clusters for longer and slows down the destruction.
\begin{table}[]
    \centering
    \caption{Comparison of process timescales at different distances.}
    \begin{tabular}{l l l l l}
    \hline
        $r$ [au] & $\tau_\text{H}$ [s] & $\tau_\text{clu}$ [yr] & $\tau_\text{ad}$ [yr] & $\tau_\text{X-ray}$ [yr] \\\hline \hline
         1 & $3.5 \cdot 10^{-4}$ & $5.4\cdot10^{-3}$ & $3.4\cdot 10^{-1}$ & $1.8 \cdot 10^{-1}$\\
         10 & $3.5 \cdot 10^{-1}$ & $5.4\cdot 10^{0}$ & $3.4\cdot 10^{2}$ & $1.8\cdot 10^{1}$   \\
         100 & $3.5 \cdot 10^{2}$ & $5.4\cdot 10^{3}$  & $3.4\cdot 10^{5}$ & $1.8\cdot 10^{3}$  \\
         1000 & $3.5 \cdot 10^{5}$ & $5.4\cdot 10^{6}$ & $3.4\cdot 10^{8}$ & $1.8\cdot 10^{5}$  \\
         \hline     
    \end{tabular}
    \tablefoot{Notes: Typical rehydrogenation timescale $k_\text{H}$, clustering timescale $k_\text{clu}$, adsorption timescale $k_\text{ad}$, and X-ray absorption timescale for coronene in the photosphere of a classical T\,Tauri disc (at three disc scale heights). At all distances, rehydrogenation is faster than the absorption of another X-ray, so a PAH is likely to always be completely hydrogenated before it absorbs another X-ray. The cascade-like breakdown of adsorbed PAH clusters by X-rays is always faster than the rebuilding processes in the outer disc, whereas in the inner disc, clustering and adsorption   compete against X-ray destruction.}
    \label{tab:rates}
\end{table}

Hence, the gas-phase abundance is enhanced compared to small PAH species for longer exposure times.\\
\\We conclude that the resulting abundance from this process is much lower than the abundance of PAHs derived from T\,Tauri observations where the 11.2\,$\mu$m feature has been detected (0.01 - 0.1 times ISM), even though much lower abundances are expected in discs without detections \citep{Geers2006}.
Therefore, as suggested by \citet{Siebenmorgen2010}, vertical mixing may be able to supply the photosphere with unprocessed PAH-carrying grains from which PAHs have not been desorbed yet.
Hence, our calculations only provide estimates for a non-mixing disc and lower estimates for a mixing disc for which the PAH abundance must be calculated using a model that considers vertical mixing.
Future observations with JWST, which has a higher sensitivity than Spitzer, will either provide us with new abundances or will establish new upper limits for the abundance of PAHs in discs. Thus, without a replenishment of PAHs from the midplane, we do not expect significant new detections of PAHs in T\,Tauri discs.

\subsubsection*{Rebuilding PAH molecules and clusters: Rehydrogenation, clustering, and adsorption}
In our dissociation cascade, we assumed that the timescales for the counter-processes of desorption and cluster dissociation are slower than the timescale of X-ray absorption.
In the following, we want to estimate under what conditions this approach is valid by calculating the timescales for rehydrogenation, clustering, and adsorption.
For each of these processes the rate can be estimated from the frequency of collisions between the corresponding reactants:
\begin{equation}
Z_\text{a,b} / n_\text{a} = k_\text{a,b} = n_\text{b} \sigma_\text{a,b} \Delta v_\text{a,b}  \text{.}
\end{equation}
Here $Z_\text{a,b}$ denotes the collision frequency between collision partners a and b, $n_\text{a}$ and $n_\text{b}$ are the number densities for a and b, and $\sigma_\text{a,b}$ describes their geometrical collision cross-section.
Furthermore, $\Delta v_\text{a,b}$ describes their relative velocity assuming Brownian motion, for which we assume a reference temperature of 400\,K.
For the rehydrogenation of coronene, we estimate a rehydrogenation rate of
\begin{equation}
    k_\text{H} =  8.9 \cdot 10^{-10} \text{cm$^3$/s} \cdot n_\text{H} \cdot\sqrt{\frac{T}{400\,\text{K}}} \,\text{.}
\end{equation}
To estimate the clustering timescale, we estimate the timescale for a collision between a monomer and another monomer where the radius of a PAH is similarly described as in equation \eqref{eq:surfaceareaPAH}.
Then, assuming that at maximum 1/100 ($t=0$ in figure \ref{fig:fraction_all}) of the PAHs are in the gas phase, we find
\begin{equation}
    k_\text{clu} = 3.6 \cdot 10^{-18} \text{cm$^3$/s} \cdot n_\text{H} \cdot\sqrt{\frac{T}{400\,\text{K}}} \,\text{.}
\end{equation}
In order to estimate the adsorption timescale, we need to make assumptions about the dust grain population.
We assumed a dust-to-gas ratio of 1:100 for spherical compact dust grains with an average density of $\rho_\text{d}=3.6$\,g/cm$^3$.
The dust grain population follows a power law of $n(a)\propto a^{-3.5}$ with $a_\text{min}=0.1$\,$\mu$m and $a_\text{max} = 0.1$\,cm and is assumed to be well mixed within the disc.
Then, by integrating the collision cross-sections with every dust grain size, we find
\begin{equation}
    k_\text{ad} = 5.8 \cdot 10^{-20} \text{cm$^3$/s} \cdot n_\text{H} \cdot\sqrt{\frac{T}{400\,\text{K}}} \,\text{.}
\end{equation}
Finally, assuming a typical disc following a surface density profile derived from the minimum mass solar nebular (MMSN) \citep{Weidenschilling1977} $\Sigma(r) = 730\,\text{g/cm$^3$} (r/\text{au})^{-1.5} (M_*/M_\odot)$ and a typical temperature profile \citep{Hayashi1981} $T(r) = 280\,\text{K} (r/\text{au})^{-0.5}(L_*/L_\odot)^{0.25}$, we can derive the corresponding rates for a typical disc around a star such as BP\,Tau.
The calculated rates are presented in Table \ref{tab:rates}.\\
\\In summary, rehydrogenation occurs on shorter timescales  than X-ray absorption, so all PAHs are expected to be fully hydrogenated before another X-ray is absorbed.
The estimations for clustering and adsorption show that in the outer disc regions, the dissociation cascade of PAHs through X-rays is faster than the rebuilding processes. In the inner disc ($r\leq30$\,au) clustering and adsorption compete with X-ray destruction. Our model prediction (no competition) needs to be treated as an upper limit for destruction, as some monomers might partially  build back clusters that can protect them from total destruction of the PAHs.
A disc with lower X-ray fluxes than our assumed model (e.g. through partial shielding) is likely to preserve even more PAHs.

\subsubsection*{Survival of PAHs in the Solar System}
Our aim was to relate our study to the amount of PAHs that can be
found in the Solar System.  It has been supposed that the Sun was a
highly active T\,Tauri star (see \citet{Güdel2007} for a review), so
the circumstellar disc of the early Solar System was intensely
irradiated with X-rays.  Today, PAHs can be found at high
concentrations close to chondrules in the Allende and Murchison
meteorites \citep{Plows2003}, the Martian meteorite ALH84001
\citep{Clemett1998, Becker1999}, and in the atmosphere of Titan
\citep{Lopez-Puertas2013}.  However, reaction pathways in Titan's cold
atmosphere have been proposed to produce PAHs locally
\citep{Zhao2018}.  PAHs have also been found in captured
interplanetary dust particles (IDPs) \citep{Clemett1993} and in
comet 81P/Wild 2 as part of the stardust mission \citep{Sandford2006}.
Tentatively, PAHs might be the carriers of the 3.28\,$\mu$m feature
\citep{Bockelee-Morvan1995} observed in Halley's Comet
\citep{Baas1986}, and Comet Levy \citep{Davies1991}, among others.
The presence of PAHs to this day in the Solar System indicates that
the primordial PAHs must have persisted throughout the protoplanetary
disc stage with heavy X-ray bombardment and beyond, or that,
alternatively, PAHs must have been formed on or inside planetesimals
through chemical reactions.  However, the low abundance of PAHs
indicates that most of them must have been destroyed or chemically
processed.  Other processes such as sooting and burning with oxygen in
the gas phase and on the surfaces of grains should be considered when
evaluating the evolution of PAHs in the Solar System \citep{Kress2010,
  Anderson2017}.

\subsection{Comparison to laboratory work}
\begin{figure}
    \vspace{-0.55cm}
    \centering 
    \includegraphics[width=0.99\linewidth]{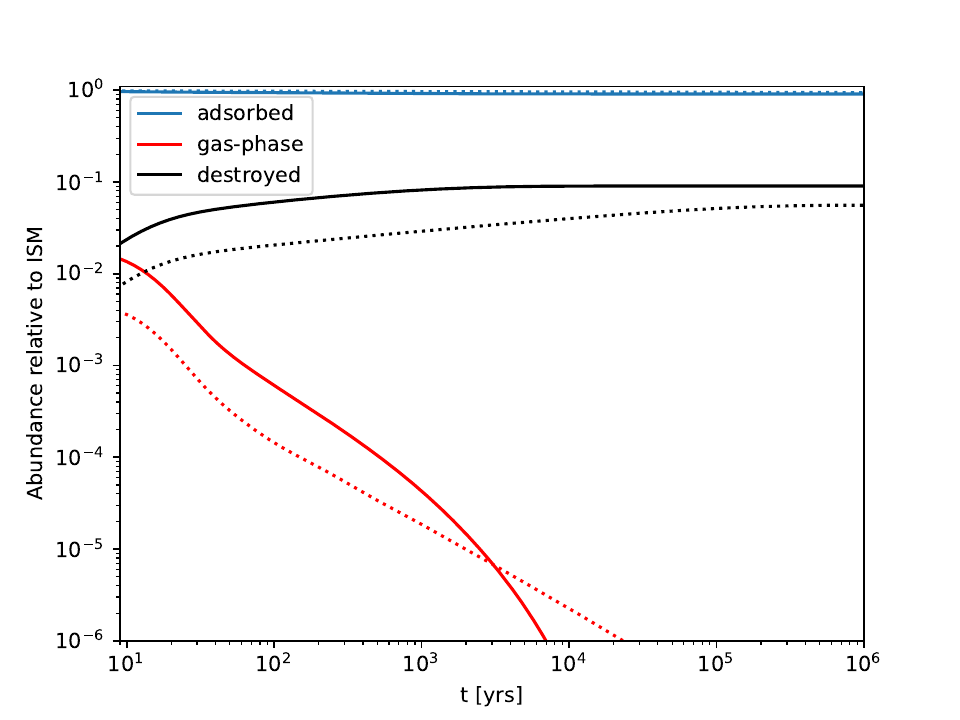}
    \caption{Summed abundance for all cluster sizes (coronene: \textit{solid lines}, ovalene: \textit{dotted lines}) relative to the ISM. All clusters that desorbed, dissociated, and then experienced \ce{C2H2} loss are marked in \textit{black}. The resulting gas-phase monomer fraction resulting from the desorption and subsequent dissociation by X-rays is shown in \textit{red}. Since most PAHs are adsorbed in the form of clusters with more than 25 molecules, the adsorbed fraction of PAHs hardly changes at all.}
    \label{fig:fraction_all}
    \vspace{-0.4cm}
\end{figure}
The absorption of X-rays has also been studied in the laboratory by several groups.
\citet{Huo2022} studied the destruction of normal- and super-hydrogenated coronene cations in an ion trap under irradiation of photons with energies from 280-300\,eV.
They found that when irradiated with X-rays whose energy is near the K edge of carbon, 30\,\% of PAHs suffer carbon loss.
This number increases from 30\,\% to 70\,\% when X-rays with energies above the K edge of carbon are considered.
Other laboratory work has been done by \citet{Monfredini2019} who exposed naphthalene (\ce{C10H8}), anthracene (\ce{C14H10}), and pyrene (\ce{C16H10}) to 2.5\,keV X-rays.
In their study under these harsh conditions, only 6.6-14.8\,\% of the PAHs survive without any carbon loss.\\
\\ Comparing these numbers to our dissociation model of single gas-phase PAHs, we found that we are underestimating the fraction of PAHs where the carbon backbone is damaged.
One plausible explanation is the potential underestimation of the deposited vibrational energy because higher energies also (slightly) increase the dissociation probability of acetylene (\ce{C2H2}) in our model.
Another possible explanation can be attributed to the fact that we have only considered the loss of acetylene (\ce{C2H2}) as a carbon loss channel, whereas fragmentation into other aliphatic carbon structures was not considered.
It seems plausible that acetylene (\ce{C2H2}) cannot be the only fragment ejected from the PAHs as fragmentation spectra show distinguished peaks at all possible carbon fragment sizes smaller than the parent molecule with a varying number of hydrogen atoms attached to the carbon fragments \citep{Monfredini2019}.
Hence, the consideration of other fragmentation channels would likely increase the probability of the PAH suffering from carbon loss in our model.
Unfortunately, these dissociation channels are seldom described in the literature from laboratory studies.
One possibility is the loss of (\ce{C4H2}) groups, which is possible under the right hydrogenation conditions; another is the weakening of the carbon skeleton by aliphatic bonds \citep{Tang2022}.\\
\\Another critical aspect of our chemical model is the uncertainties of the activation energy $E_\text{A}$ and the entropy factor $\Delta S$ of the reactions.
The dissociation rate depends exponentially on both values so that small variations in $E_\text{A}$ and $\Delta S$ can change the dissociation rate by several orders of magnitude.
As a comparison between the dissociation parameters obtained by \citet{Ling1995} to a more recent study by \citet{West2018} shows, the structure of the PAH and the  model approach used strongly affect the derived activation energies $E_\text{A}$ and entropy factors $\Delta S$ for certain reactions.
The given uncertainties (e.g. choosing the values for the release of hydrogen from \citet{West2018} for pyrene \ce{C16H10} with $E_\text{A} =4.16 \pm 0.69$\,eV and $\Delta S = -5 \pm 37$) result in a change of dissociation rates of multiple orders of magnitude.
Additionally, \citet{Castellanos2018b} showed that hydrogen roaming along the carbon rings needs to be considered in interpreting laboratory data and modelling hydrogen loss.
Hence, the probability of each event in our model and the corresponding cascade model strongly depends on the considered dissociation model.
Therefore, we emphasise the need for detailed chemical studies of the underlying fragmentation channels of \ce{C2H2} for PAHs similar to the study of \citet{Castellanos2018b}.\\
\\If coronene is superhydrogenated, the chance of carbon loss is significantly decreased to around 40\,\% for X-rays carrying more than 280\,eV energy \citet{Huo2022}.
Therefore, the attachment of additional hydrogen atoms to the PAH acts as an additional mechanism to stabilise the carbon backbone against hard radiation, which was also concluded by \citet{Reitsma2014}.
The authors studied the fragmentation channels of superhydrogenated coronene (\ce{C24H12}) with up to seven additional hydrogen atoms.
In a separate theoretical study, \citet{Andrews2016} calculated the fraction of superhydrogenated PAHs for standard ISM conditions.
Their results show that in order to have a significant fraction of abundant superhydrogenated PAHs (depending on the PAH size $\approx 20\,\%$ for coronene, 100\,\% for circumcoronene), $G_0/n_\text{H} < 10^{-3}$ is required.
Hence in typical T\,Tauri disc conditions, superhydrogenation is possible as $G_0 \approx 10^4$ at 1\,au, which could slow down the destruction of the carbon backbone of the PAHs in protoplanetary discs.
However, coronene with seven additional hydrogen atoms has only 16.5\,eV of binding energy available for cooling by de-hydrogenation of the additional hydrogen \citep{Reitsma2014}. 
It is unclear whether superhydrogenation can stop the destruction of PAHs by X-ray photon absorption entirely.
A detailed chemical model is necessary to fully characterise the hydrogenation and charge state of the PAHs in protoplanetary discs to determine their increased stability against hard radiation.

\subsection{Comparison to theoretical work by Siebenmorgen}
Older theoretical work done by \citet{Siebenmorgen2010} and \citet{Siebenmorgen2012} has investigated the destruction of PAHs through high-energy radiation in T\,Tauri discs.
In these studies, the authors examine the destruction of individual PAH molecules through the absorption of high-energy photons, assuming that the X-ray and EUV photons transfer all of their energy to the molecule.
In these models, the absorption of a single X-ray photon with an energy of 1\,keV has the potential to completely disintegrate a PAH consisting of 100 carbon atoms into individual atoms.
However, the ejected primary electron of the ionisation process carries away the remaining energy of the X-ray as kinetic energy.
Hence, the amount of energy that can be transferred into vibrational excitation to trigger molecular dissociation is much lower than the energy of an X-ray photon.
The introduction of H and H$_2$ loss  opens new energy dissipation channels in our model so that not every X-ray can destroy a PAH monomer (see Fig. \ref{fig:gasphasemonomer}).
This leads to destruction timescales of PAHs that are a factor of  3-5 longer than in the model by Siebenmorgen et al.
Nevertheless, our gas-phase PAH destruction timescales are still very short compared to the lifetime of a protoplanetary disc.\\
\\The authors relate the survivability of PAHs to the strength of turbulence in a disc. 
By vertical mixing into the radiation shielded midplane, the time an individual PAH is exposed to hard radiation is limited to a timescale much shorter than the lifetime of a disc, and the midplane can act as a reservoir to replenish destroyed PAHs.
In relation to our calculations, vertical mixing is also an effective mechanism to retain a higher PAH abundance that delivers new dust grains with adsorbed small clusters into the photosphere that are desorbed and destroyed by X-rays.
Therefore, we expect the decreasing fraction of gas-phase PAHs in Figure \ref{fig:fraction_all} as a lower limit of the PAH abundance if vertical mixing is active in the disc.

\subsection{PAH destruction supporting organic chemistry}
It is unknown what contribution PAHs make to the organic chemistry of discs and what their photochemical products look like.
The destruction of  gas-phase PAHs through X-rays could contribute to the formation of more complex species such as fullerenes \ce{C60} with spectral features at 17.4 and 18.9\,$\mu$m \citep{Cami2010, Sellgren2010}.
High temperatures and strong excitation support the top-down formation of fullerenes \citep{Berne2012, Zhen2014}. This has been investigated in the reflection nebula NGC 7032 driven by UV photons \citep{Berne2015}.
However, a possible build-up or repair path for PAHs after carbon loss might be through reactions with \ce{C2H2} at warm temperatures where warm carbon chemistry can act \citep{Carr2008, Carr2011, Bast2013}.
An investigation of a possible link between PAHs and the rarely seen nanodiamonds \citep{Acke2006} with features at 3.43 and 3.53\,$\mu$m would also be interesting to study.\\
\\For completeness, we would like to mention the recently observed protoplanetary disc (2MASS-J16053215-1933159) that has a highly enhanced C/O ratio with detections of \ce{C2H2}, \ce{C6H6}, and other organic compounds.
PAH signatures and small silicate grains are completely absent in this system \citep{Tabone2023}.
Even though this disc is rather exceptional, we would like to mention the lack of PAHs and the high abundance of carbon-bearing molecules in this system.

\section{Conclusions}
\label{sec:conclusions}
In this exploratory work we modelled the interaction of PAHs, PAH clusters, and adsorbed PAH clusters with X-rays.
The X-rays   vibrationally excite the PAH by $15-35$\,eV and doubly ionise it if the X-ray energy is more than the energy barrier of the carbon K edge of 280\,eV.
Subsequently, the highly excited vibrational state can
do the following:
\begin{itemize}
    \item cause desorption of clusters from dust grains up to a size of 25 cluster members for coronene \ce{c24H12} and two members for circumcoronene \ce{C54H18}. The size of the PAH species strongly limits the largest desorbable cluster size within the lifetime of a protoplanetary disc. After desorption, small clusters carry enough energy to break down without additional excitation.
    \item efficiently break down PAH clusters into monomers. If a cluster can desorb from a grain surface, two or fewer additional X-rays can break the entire coronene cluster into monomers. However based on the presented timescales of each process, partial rebuilding of a cluster is possible in the inner parts of the disc  ($r\leq30$\,au)  that can protect PAHs from complete destruction. Therefore, our models predict upper limits for the destruction of PAHs.
    For larger clusters, the dissociation of a cluster is less efficient and a large number of X-rays is required to break down a coronene cluster with 60 cluster members \ce{(C24H12)60}.
    \item cause strong dehydrogenation and damage to the carbon structure of the PAH by the loss of acetylene \ce{C2H2}. The destruction of monomers is independent of the considered PAH species, and acetylene loss occurs after three X-ray absorption timescales. However, the different carbon-loss channels for coronene and larger PAHs are not well understood. We therefore emphasise the need to study the carbon fragmentation of highly excited PAHs and their mechanisms in chemical studies.
\end{itemize}
As a result, only a small fraction of PAHs can leave the grain surface and enter the gas phase as monomers.
Our expected PAH abundance for T\,Tauri discs decreased by a factor of 100 or higher compared to the ISM abundance and will decrease over time.
Hence, in discs in which a higher abundance is measured, a replenishing mechanism must act to support the photosphere of the disc with unprocessed PAHs, or another mechanism must evaporate more PAHs from the dust grain surface.\\
\\Finally, we are curious about the latest observations using JWST.
Our models predict a significantly lower PAH abundance than JWST could observe if vertical transport is not effective.
Hence, if many new detections of PAHs occur in T\,Tauri discs in the future, this would allow us to draw conclusions about the vertical mixing in some of these discs.
We emphasise the need to approach the chemistry of  PAHs  and their transport in complex disc and chemistry models in order to interpret future results from observations.

\begin{acknowledgements}
The authors thank Alessandra Candian for very interesting and useful discussions that improved this manuscript, and her insights in molecular dynamics of PAHs. The authors also thank the anonymous referee for valuable input and constructive criticism.
K.L. acknowledges funding from the Nederlandse Onderzoekschool Voor Astronomie (NOVA) project number R.2320.0130. C.D. acknowledges funding from the Netherlands Organisation for Scientific Research (NWO) TOP-1 grant as part of the research program “Herbig Ae/Be stars, Rosetta stones for understanding the formation of planetary systems”, project number 614.001.552. 
Studies of interstellar PAHs at Leiden Observatory are supported through a Spinoza award from the Dutch research council, NWO.
\end{acknowledgements}
\vspace{-0.6cm}

\bibliographystyle{aa} 
\bibliography{library.bib} 

\begin{thebibliography}{84}
\expandafter\ifx\csname natexlab\endcsname\relax\def\natexlab#1{#1}\fi

\bibitem[{{Acke} {et~al.}(2010){Acke}, {Bouwman}, {Juh{\'a}sz}, {Henning}, {van
  den Ancker}, {Meeus}, {Tielens}, \& {Waters}}]{Acke2010}
{Acke}, B., {Bouwman}, J., {Juh{\'a}sz}, A., {et~al.} 2010, \apj, 718, 558

\bibitem[{{Acke} \& {van den Ancker}(2004)}]{Acke2004}
{Acke}, B. \& {van den Ancker}, M.~E. 2004, \aap, 426, 151

\bibitem[{{Acke} \& {van den Ancker}(2006)}]{Acke2006}
{Acke}, B. \& {van den Ancker}, M.~E. 2006, \aap, 457, 171

\bibitem[{{Allamandola} {et~al.}(1985){Allamandola}, {Tielens}, \&
  {Barker}}]{Allamandola1985}
{Allamandola}, L.~J., {Tielens}, A.~G.~G.~M., \& {Barker}, J.~R. 1985, \apjl,
  290, L25

\bibitem[{{Allamandola} {et~al.}(1989){Allamandola}, {Tielens}, \&
  {Barker}}]{Allamandola1989}
{Allamandola}, L.~J., {Tielens}, A.~G.~G.~M., \& {Barker}, J.~R. 1989, \apjs,
  71, 733

\bibitem[{{Alofi} \& {Srivastava}(2014)}]{Alofi2014}
{Alofi}, A. \& {Srivastava}, G.~P. 2014, Applied Physics Letters, 104, 031903

\bibitem[{{Anderson} {et~al.}(2017){Anderson}, {Bergin}, {Blake}, {Ciesla},
  {Visser}, \& {Lee}}]{Anderson2017}
{Anderson}, D.~E., {Bergin}, E.~A., {Blake}, G.~A., {et~al.} 2017, \apj, 845,
  13

\bibitem[{{Andrews} {et~al.}(2016){Andrews}, {Candian}, \&
  {Tielens}}]{Andrews2016}
{Andrews}, H., {Candian}, A., \& {Tielens}, A.~G.~G.~M. 2016, \aap, 595, A23

\bibitem[{{Appenzeller} \& {Mundt}(1989)}]{Appenzeller1989}
{Appenzeller}, I. \& {Mundt}, R. 1989, \aapr, 1, 291

\bibitem[{Auger(1923)}]{Auger1923}
Auger, P. 1923, C.R.A.S., 177, 169

\bibitem[{{Baas} {et~al.}(1986){Baas}, {Geballe}, \& {Walther}}]{Baas1986}
{Baas}, F., {Geballe}, T.~R., \& {Walther}, D.~M. 1986, \apjl, 311, L97

\bibitem[{{Bakes} {et~al.}(2001){Bakes}, {Tielens}, \&
  {Bauschlicher}}]{Bakes2001}
{Bakes}, E.~L.~O., {Tielens}, A.~G.~G.~M., \& {Bauschlicher}, Charles~W., J.
  2001, \apj, 556, 501

\bibitem[{Bambynek {et~al.}(1972)Bambynek, Crasemann, Fink, Freund, Mark,
  Swift, Price, \& Rao}]{Bambynek1972}
Bambynek, W., Crasemann, B., Fink, R.~W., {et~al.} 1972, Rev. Mod. Phys., 44,
  716

\bibitem[{{Bast} {et~al.}(2013){Bast}, {Lahuis}, {van Dishoeck}, \&
  {Tielens}}]{Bast2013}
{Bast}, J.~E., {Lahuis}, F., {van Dishoeck}, E.~F., \& {Tielens}, A.~G.~G.~M.
  2013, \aap, 551, A118

\bibitem[{{Becker} {et~al.}(1999){Becker}, {Popp}, {Rust}, \&
  {Bada}}]{Becker1999}
{Becker}, L., {Popp}, B., {Rust}, T., \& {Bada}, J.~L. 1999, Earth and
  Planetary Science Letters, 167, 71

\bibitem[{{Bern{\'e}} {et~al.}(2015){Bern{\'e}}, {Montillaud}, \&
  {Joblin}}]{Berne2015}
{Bern{\'e}}, O., {Montillaud}, J., \& {Joblin}, C. 2015, \aap, 577, A133

\bibitem[{{Bern{\'e}} \& {Tielens}(2012)}]{Berne2012}
{Bern{\'e}}, O. \& {Tielens}, A. G.~G.~M. 2012, Proceedings of the National
  Academy of Science, 109, 401

\bibitem[{Biggs \& Lighthill(1988)}]{Biggs1988}
Biggs, F. \& Lighthill, R. 1988, Sandia Report

\bibitem[{{Bockel{\'e}e-Morvan} {et~al.}(1995){Bockel{\'e}e-Morvan}, {Brooke},
  \& {Crovisier}}]{Bockelee-Morvan1995}
{Bockel{\'e}e-Morvan}, D., {Brooke}, T.~Y., \& {Crovisier}, J. 1995, \icarus,
  116, 18

\bibitem[{{Boechat-Roberty} {et~al.}(2009){Boechat-Roberty}, {Neves},
  {Pilling}, {Lago}, \& {de Souza}}]{Boechat-Roberty2009}
{Boechat-Roberty}, H.~M., {Neves}, R., {Pilling}, S., {Lago}, A.~F., \& {de
  Souza}, G.~G.~B. 2009, \mnras, 394, 810

\bibitem[{{Bouwman} {et~al.}(2011{\natexlab{a}}){Bouwman}, {Cuppen},
  {Steglich}, {Allamandola}, \& {Linnartz}}]{Bouwman2011b}
{Bouwman}, J., {Cuppen}, H.~M., {Steglich}, M., {Allamandola}, L.~J., \&
  {Linnartz}, H. 2011{\natexlab{a}}, \aap, 529, A46

\bibitem[{{Bouwman} {et~al.}(2011{\natexlab{b}}){Bouwman}, {Mattioda},
  {Linnartz}, \& {Allamandola}}]{Bouwman2011}
{Bouwman}, J., {Mattioda}, A.~L., {Linnartz}, H., \& {Allamandola}, L.~J.
  2011{\natexlab{b}}, \aap, 525, A93

\bibitem[{Brittain {et~al.}(2023)Brittain, Kamp, Meeus, Oudmaijer, \&
  Waters}]{Brittain2023}
Brittain, S.~D., Kamp, I., Meeus, G., Oudmaijer, R.~D., \& Waters, L. B. F.~M.
  2023, Herbig Stars: A Quarter Century of Progress

\bibitem[{{Cami} {et~al.}(2010){Cami}, {Bernard-Salas}, {Peeters}, \&
  {Malek}}]{Cami2010}
{Cami}, J., {Bernard-Salas}, J., {Peeters}, E., \& {Malek}, S.~E. 2010,
  Science, 329, 1180

\bibitem[{{Carr} \& {Najita}(2008)}]{Carr2008}
{Carr}, J.~S. \& {Najita}, J.~R. 2008, Science, 319, 1504

\bibitem[{{Carr} \& {Najita}(2011)}]{Carr2011}
{Carr}, J.~S. \& {Najita}, J.~R. 2011, \apj, 733, 102

\bibitem[{{Castellanos} {et~al.}(2018){Castellanos}, {Candian}, {Zhen},
  {Linnartz}, \& {Tielens}}]{Castellanos2018b}
{Castellanos}, P., {Candian}, A., {Zhen}, J., {Linnartz}, H., \& {Tielens},
  A.~G.~G.~M. 2018, \aap, 616, A166

\bibitem[{{Clemett} {et~al.}(1998){Clemett}, {Dulay}, {Gillette}, {Chillier},
  {Mahajan}, \& {Zare}}]{Clemett1998}
{Clemett}, S.~J., {Dulay}, M.~T., {Gillette}, J.~S., {et~al.} 1998, Faraday
  Discussions, 109, 417

\bibitem[{{Clemett} {et~al.}(1993){Clemett}, {Maechling}, {Zare}, {Swan}, \&
  {Walker}}]{Clemett1993}
{Clemett}, S.~J., {Maechling}, C.~R., {Zare}, R.~N., {Swan}, P.~D., \&
  {Walker}, R.~M. 1993, Science, 262, 721

\bibitem[{{Davies} {et~al.}(1991){Davies}, {Green}, \& {Geballe}}]{Davies1991}
{Davies}, J.~K., {Green}, S.~F., \& {Geballe}, T.~R. 1991, \mnras, 251, 148

\bibitem[{de~Andres {et~al.}(2008)de~Andres, Ram\'{\i}rez, \&
  Verg\'es}]{Andres2008}
de~Andres, P.~L., Ram\'{\i}rez, R., \& Verg\'es, J.~A. 2008, Phys. Rev. B, 77,
  045403

\bibitem[{{Desert} \& {Dennefeld}(1988)}]{Desert1988}
{Desert}, F.~X. \& {Dennefeld}, M. 1988, \aap, 206, 227

\bibitem[{{Draine} \& {Li}(2007)}]{Draine2007}
{Draine}, B.~T. \& {Li}, A. 2007, \apj, 657, 810

\bibitem[{{Duley} \& {Williams}(1981)}]{Duley1981}
{Duley}, W.~W. \& {Williams}, D.~A. 1981, \mnras, 196, 269

\bibitem[{{Geers} {et~al.}(2006){Geers}, {Augereau}, {Pontoppidan},
  {Dullemond}, {Visser}, {Kessler-Silacci}, {Evans}, {van Dishoeck}, {Blake},
  {Boogert}, {Brown}, {Lahuis}, \& {Mer{\'\i}n}}]{Geers2006}
{Geers}, V.~C., {Augereau}, J.~C., {Pontoppidan}, K.~M., {et~al.} 2006, \aap,
  459, 545

\bibitem[{{Geers} {et~al.}(2009){Geers}, {van Dishoeck}, {Pontoppidan},
  {Lahuis}, {Crapsi}, {Dullemond}, \& {Blake}}]{Geers2009}
{Geers}, V.~C., {van Dishoeck}, E.~F., {Pontoppidan}, K.~M., {et~al.} 2009,
  \aap, 495, 837

\bibitem[{{Grankin}(2016)}]{Grankin2016}
{Grankin}, K.~N. 2016, Astronomy Letters, 42, 314

\bibitem[{G{\"u}del(2007)}]{Güdel2007}
G{\"u}del, M. 2007, Living Reviews in Solar Physics, 4, 1

\bibitem[{{Hayashi}(1981)}]{Hayashi1981}
{Hayashi}, C. 1981, Progress of Theoretical Physics Supplement, 70, 35

\bibitem[{{Huo} {et~al.}(2022){Huo}, {Cangahuala}, {Goulart}, {Zamudio-Bayer},
  {Kubin}, {Timm}, {Lau}, {von Issendorff}, {Hoekstra}, {Faraji}, \&
  {Schlath{\"o}lter}}]{Huo2022}
{Huo}, Y., {Cangahuala}, M. K.~E., {Goulart}, M., {et~al.} 2022, \pra, 106,
  063104

\bibitem[{Jayadev {et~al.}(2023)Jayadev, Ferino-Pérez, Matz, Krylov, \&
  Jagau}]{Jayadev2023}
Jayadev, N.~K., Ferino-Pérez, A., Matz, F., Krylov, A.~I., \& Jagau, T.-C.
  2023, The Journal of Chemical Physics, 158, 064109

\bibitem[{{Jochims} {et~al.}(1994){Jochims}, {Ruhl}, {Baumgartel}, {Tobita}, \&
  {Leach}}]{Jochims1994}
{Jochims}, H.~W., {Ruhl}, E., {Baumgartel}, H., {Tobita}, S., \& {Leach}, S.
  1994, \apj, 420, 307

\bibitem[{{Kamp}(2011)}]{Kamp2011}
{Kamp}, I. 2011, in EAS Publications Series, Vol.~46, EAS Publications Series,
  ed. C.~{Joblin} \& A.~G.~G.~M. {Tielens}, 271--283

\bibitem[{{Kress} {et~al.}(2010){Kress}, {Tielens}, \& {Frenklach}}]{Kress2010}
{Kress}, M.~E., {Tielens}, A. G.~G.~M., \& {Frenklach}, M. 2010, Advances in
  Space Research, 46, 44

\bibitem[{{Lange} {et~al.}(2021){Lange}, {Dominik}, \& {Tielens}}]{Lange2021}
{Lange}, K., {Dominik}, C., \& {Tielens}, A.~G.~G.~M. 2021, \aap, 653, A21

\bibitem[{{Lange} {et~al.}(2023){Lange}, {Dominik}, \& {Tielens}}]{Lange2023}
{Lange}, K., {Dominik}, C., \& {Tielens}, A.~G.~G.~M. 2023, \aap, 674, A200

\bibitem[{{Leger} \& {Puget}(1984)}]{Leger1984}
{Leger}, A. \& {Puget}, J.~L. 1984, \aap, 137, L5

\bibitem[{{Leger} \& {Rouan}(1986)}]{Leger1986}
{Leger}, A. \& {Rouan}, D. 1986, \aap, 162, 211

\bibitem[{{Li}(2009)}]{Li2009}
{Li}, A. 2009, in Deep Impact as a World Observatory Event: Synergies in Space,
  Time, and Wavelength, ed. H.~U. {K{\"a}ufl} \& C.~{Sterken}, 161

\bibitem[{{Li}(2020)}]{Li2020}
{Li}, A. 2020, Nature Astronomy, 4, 339

\bibitem[{Ling {et~al.}(1995)Ling, Gotkis, \& Lifshitz}]{Ling1995}
Ling, Y., Gotkis, Y., \& Lifshitz, C. 1995, European Mass Spectrometry, 1, 41

\bibitem[{Ling \& Lifshitz(1998)}]{Ling1998}
Ling, Y. \& Lifshitz, C. 1998, The Journal of Physical Chemistry A, 102, 708

\bibitem[{{L{\'o}pez-Puertas} {et~al.}(2013){L{\'o}pez-Puertas}, {Dinelli},
  {Adriani}, {Funke}, {Garc{\'\i}a-Comas}, {Moriconi}, {D'Aversa}, {Boersma},
  \& {Allamandola}}]{Lopez-Puertas2013}
{L{\'o}pez-Puertas}, M., {Dinelli}, B.~M., {Adriani}, A., {et~al.} 2013, \apj,
  770, 132

\bibitem[{{Micelotta} {et~al.}(2010){Micelotta}, {Jones}, \&
  {Tielens}}]{Micelotta2010}
{Micelotta}, E.~R., {Jones}, A.~P., \& {Tielens}, A.~G.~G.~M. 2010, \aap, 510,
  A37

\bibitem[{{Monfredini} {et~al.}(2019){Monfredini}, {Quiti{\'a}n-Lara},
  {Fantuzzi}, {Wolff}, {Mendoza}, {Lago}, {Sales}, {Pastoriza}, \&
  {Boechat-Roberty}}]{Monfredini2019}
{Monfredini}, T., {Quiti{\'a}n-Lara}, H.~M., {Fantuzzi}, F., {et~al.} 2019,
  \mnras, 488, 451

\bibitem[{{Pallavicini} {et~al.}(1981){Pallavicini}, {Golub}, {Rosner},
  {Vaiana}, {Ayres}, \& {Linsky}}]{Pallavicini1981}
{Pallavicini}, R., {Golub}, L., {Rosner}, R., {et~al.} 1981, \apj, 248, 279

\bibitem[{{Peeters} {et~al.}(2002){Peeters}, {Mart{\'\i}n-Hern{\'a}ndez},
  {Damour}, {Cox}, {Roelfsema}, {Baluteau}, {Tielens}, {Churchwell}, {Kessler},
  {Mathis}, {Morisset}, \& {Schaerer}}]{Peeters2002}
{Peeters}, E., {Mart{\'\i}n-Hern{\'a}ndez}, N.~L., {Damour}, F., {et~al.} 2002,
  \aap, 381, 571

\bibitem[{{Plows} {et~al.}(2003){Plows}, {Elsila}, {Zare}, \&
  {Buseck}}]{Plows2003}
{Plows}, F.~L., {Elsila}, J.~E., {Zare}, R.~N., \& {Buseck}, P.~R. 2003, \gca,
  67, 1429

\bibitem[{{Preibisch} \& {Feigelson}(2005)}]{Preibisch2005b}
{Preibisch}, T. \& {Feigelson}, E.~D. 2005, \apjs, 160, 390

\bibitem[{{Preibisch} {et~al.}(2005){Preibisch}, {McCaughrean}, {Grosso},
  {Feigelson}, {Flaccomio}, {Getman}, {Hillenbrand}, {Meeus}, {Micela},
  {Sciortino}, \& {Stelzer}}]{Preibisch2005}
{Preibisch}, T., {McCaughrean}, M.~J., {Grosso}, N., {et~al.} 2005, \apjs, 160,
  582

\bibitem[{{Reitsma} {et~al.}(2014){Reitsma}, {Boschman}, {Deuzeman},
  {Gonz{\'a}lez-Maga{\~n}a}, {Hoekstra}, {Cazaux}, {Hoekstra}, \&
  {Schlath{\"o}lter}}]{Reitsma2014}
{Reitsma}, G., {Boschman}, L., {Deuzeman}, M.~J., {et~al.} 2014, \prl, 113,
  053002

\bibitem[{{Reitsma} {et~al.}(2015){Reitsma}, {Boschman}, {Deuzeman},
  {Hoekstra}, {Hoekstra}, \& {Schlath{\"o}lter}}]{Reitsma2015}
{Reitsma}, G., {Boschman}, L., {Deuzeman}, M.~J., {et~al.} 2015, \jcp, 142,
  024308

\bibitem[{{Russell} {et~al.}(1977){Russell}, {Soifer}, \&
  {Willner}}]{Russell1977}
{Russell}, R.~W., {Soifer}, B.~T., \& {Willner}, S.~P. 1977, \apjl, 217, L149

\bibitem[{Rye \& Houston(1984)}]{Rye1984}
Rye, R. \& Houston, J. 1984, Accounts of Chemical Research, 17, 41

\bibitem[{{Sandford} {et~al.}(2006){Sandford}, {Al{\'e}on}, {Alexander},
  {Araki}, {Bajt}, {Baratta}, {Borg}, {Bradley}, {Brownlee}, {Brucato},
  {Burchell}, {Busemann}, {Butterworth}, {Clemett}, {Cody}, {Colangeli},
  {Cooper}, {D'Hendecourt}, {Djouadi}, {Dworkin}, {Ferrini}, {Fleckenstein},
  {Flynn}, {Franchi}, {Fries}, {Gilles}, {Glavin}, {Gounelle}, {Grossemy},
  {Jacobsen}, {Keller}, {Kilcoyne}, {Leitner}, {Matrajt}, {Meibom}, {Mennella},
  {Mostefaoui}, {Nittler}, {Palumbo}, {Papanastassiou}, {Robert}, {Rotundi},
  {Snead}, {Spencer}, {Stadermann}, {Steele}, {Stephan}, {Tsou}, {Tyliszczak},
  {Westphal}, {Wirick}, {Wopenka}, {Yabuta}, {Zare}, \&
  {Zolensky}}]{Sandford2006}
{Sandford}, S.~A., {Al{\'e}on}, J., {Alexander}, C. M.~O.~D., {et~al.} 2006,
  Science, 314, 1720

\bibitem[{{Sellgren}(1984)}]{Sellgren1984}
{Sellgren}, K. 1984, \apj, 277, 623

\bibitem[{{Sellgren} {et~al.}(2010){Sellgren}, {Werner}, {Ingalls}, {Smith},
  {Carleton}, \& {Joblin}}]{Sellgren2010}
{Sellgren}, K., {Werner}, M.~W., {Ingalls}, J.~G., {et~al.} 2010, \apjl, 722,
  L54

\bibitem[{{Seok} \& {Li}(2017)}]{Seok2017}
{Seok}, J.~Y. \& {Li}, A. 2017, \apj, 835, 291

\bibitem[{{Siebenmorgen} \& {Heymann}(2012)}]{Siebenmorgen2012}
{Siebenmorgen}, R. \& {Heymann}, F. 2012, \aap, 543, A25

\bibitem[{{Siebenmorgen} \& {Kr{\"u}gel}(2010)}]{Siebenmorgen2010}
{Siebenmorgen}, R. \& {Kr{\"u}gel}, E. 2010, \aap, 511, A6

\bibitem[{{Tabone} {et~al.}(2023){Tabone}, {Bettoni}, {van Dishoeck},
  {Arabhavi}, {Grant}, {Gasman}, {Henning}, {Kamp}, {G{\"u}del}, {Lagage},
  {Ray}, {Vandenbussche}, {Abergel}, {Absil}, {Argyriou}, {Barrado},
  {Boccaletti}, {Bouwman}, {Caratti o Garatti}, {Geers}, {Glauser},
  {Justannont}, {Lahuis}, {Mueller}, {Nehm{\'e}}, {Olofsson}, {Pantin},
  {Scheithauer}, {Waelkens}, {Waters}, {Black}, {Christiaens}, {Guadarrama},
  {Morales-Calder{\'o}n}, {Jang}, {Kanwar}, {Pawellek}, {Perotti}, {Perrin},
  {Rodgers-Lee}, {Samland}, {Schreiber}, {Schwarz}, {Colina}, {{\"O}stlin}, \&
  {Wright}}]{Tabone2023}
{Tabone}, B., {Bettoni}, G., {van Dishoeck}, E.~F., {et~al.} 2023, Nature
  Astronomy

\bibitem[{{Tang} {et~al.}(2022){Tang}, {Simonsen}, {Jaganathan}, {Palot{\'a}s},
  {Oomens}, {Hornek{\ae}r}, \& {Hammer}}]{Tang2022}
{Tang}, Z., {Simonsen}, F. D.~S., {Jaganathan}, R., {et~al.} 2022, \aap, 663,
  A150

\bibitem[{{Tielens}(2008)}]{Tielens2008}
{Tielens}, A.~G.~G.~M. 2008, \araa, 46, 289

\bibitem[{{Tielens}(2013)}]{Tielens2013}
{Tielens}, A.~G.~G.~M. 2013, Reviews of Modern Physics, 85, 1021

\bibitem[{Tielens(2021)}]{Tielens2021}
Tielens, A. G. G.~M. 2021, Molecular Astrophysics (Cambridge University Press)

\bibitem[{{Valeg{\r{a}}rd} {et~al.}(2021){Valeg{\r{a}}rd}, {Waters}, \&
  {Dominik}}]{Valegard2021}
{Valeg{\r{a}}rd}, P.~G., {Waters}, L.~B.~F.~M., \& {Dominik}, C. 2021, \aap,
  652, A133

\bibitem[{{Verner} \& {Yakovlev}(1995)}]{Verner1995}
{Verner}, D.~A. \& {Yakovlev}, D.~G. 1995, \aaps, 109, 125

\bibitem[{{Verstraete} {et~al.}(1996){Verstraete}, {Puget}, {Falgarone},
  {Drapatz}, {Wright}, \& {Timmermann}}]{Verstraete1996}
{Verstraete}, L., {Puget}, J.~L., {Falgarone}, E., {et~al.} 1996, \aap, 315,
  L337

\bibitem[{{Visser} {et~al.}(2007){Visser}, {Geers}, {Dullemond}, {Augereau},
  {Pontoppidan}, \& {van Dishoeck}}]{Visser2007}
{Visser}, R., {Geers}, V.~C., {Dullemond}, C.~P., {et~al.} 2007, \aap, 466, 229

\bibitem[{{Waters} \& {Waelkens}(1998)}]{Waters1998}
{Waters}, L.~B.~F.~M. \& {Waelkens}, C. 1998, \araa, 36, 233

\bibitem[{{Weidenschilling}(1977)}]{Weidenschilling1977}
{Weidenschilling}, S.~J. 1977, \apss, 51, 153

\bibitem[{West {et~al.}(2018)West, Rodriguez~Castillo, Sit, Mohamad, Lowe,
  Joblin, Bodi, \& Mayer}]{West2018}
West, B., Rodriguez~Castillo, S., Sit, A., {et~al.} 2018, Phys. Chem. Chem.
  Phys., 20, 7195

\bibitem[{{Zhao} {et~al.}(2018){Zhao}, {Kaiser}, {Xu}, {Ablikim}, {Ahmed},
  {Evseev}, {Bashkirov}, {Azyazov}, \& {Mebel}}]{Zhao2018}
{Zhao}, L., {Kaiser}, R.~I., {Xu}, B., {et~al.} 2018, Nature Astronomy, 2, 973

\bibitem[{{Zhen} {et~al.}(2014){Zhen}, {Castellanos}, {Paardekooper},
  {Linnartz}, \& {Tielens}}]{Zhen2014}
{Zhen}, J., {Castellanos}, P., {Paardekooper}, D.~M., {Linnartz}, H., \&
  {Tielens}, A. G.~G.~M. 2014, \apjl, 797, L30

\end{thebibliography}
\begin{appendix}
\section{Probability of dissociation events}
\label{appendix:probs}
\begin{figure}
    \centering
    \vspace{-0.3cm}
    \includegraphics[width=1\linewidth]{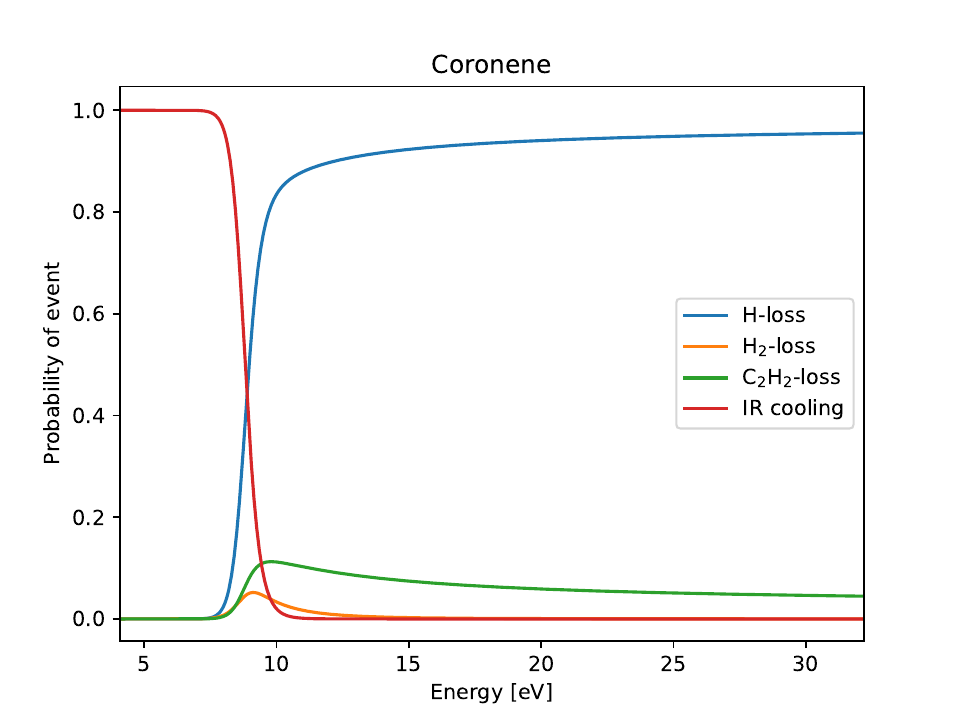}
    \caption{Probabilities for loss of H, H$_2,$ and C$_2$H$_2$ and IR cooling for coronene given an excitation energy $E$.}
    \label{fig:probabilities_zoom}
\end{figure}

The rate at which a certain event occurs given a set of rates is determined by Equation \eqref{eq:prob}. 
Using the dissociative processes for H, H$_2,$ and C$_2$H$_2$ and non-dissociative IR cooling, the probability of each depends on the excitation energy of the PAH and is displayed in Figure \ref{fig:probabilities_zoom}.
For low energies, IR cooling is dominant, while for very high excitation energies, H loss is dominant approaching a probability of 95\,\%.
The probability of acetylene loss decreases from 11\,\% at its peak at 9.8\,eV to 5\,\% for high energies.
Hence, starting at a high excitation energy, every ejected hydrogen atom slightly increases the probability of acetylene loss.
Furthermore, Figure \ref{fig:gasphasemonomer} shows a pattern that repeats every 4.6\,eV, which corresponds to the binding energy of an H atom to the PAH.
This pattern resamples the dissocation probabilities from 8.6\,eV-13.2\,eV, to which the PAH cools down before IR sets in. 
If 13.2\,eV are exceed, the PAH will lose another hydrogen and reach 8.6\,eV again.

\section{Leftover energies and probability distribution of desorbed PAH clusters}
\label{appendix:E_and_probs}
\begin{figure}[h!]
    \centering
    \includegraphics[width=1\linewidth]{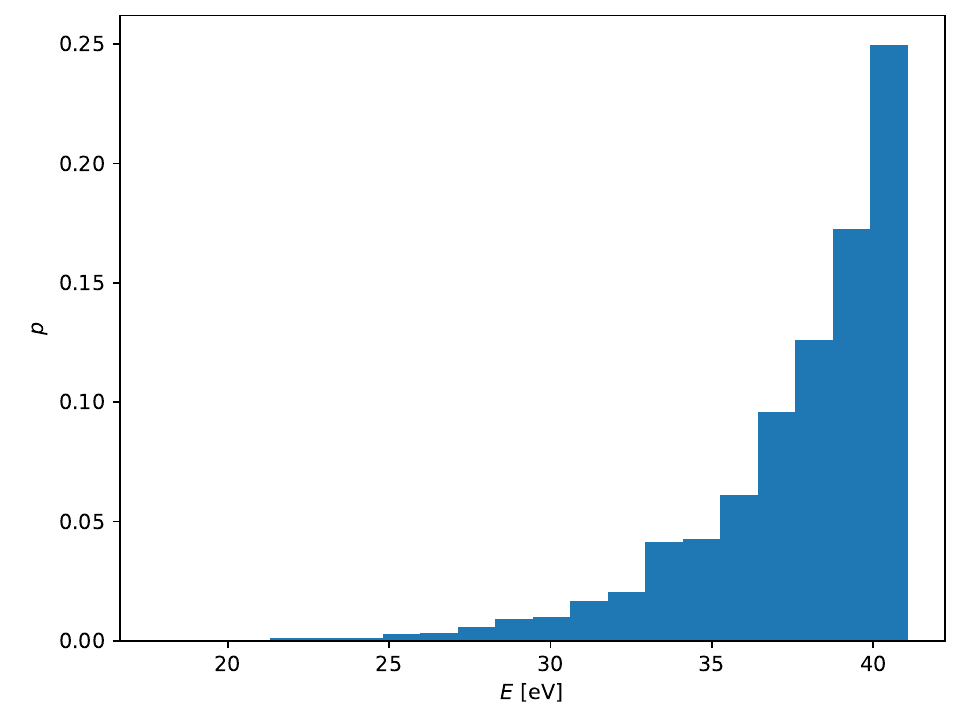}
    \caption{Histogram of leftover vibrational excitation energies after desorption of a coronene (\ce{C24H12}) decamer from a 400\,K dust grain.}
    \label{fig:E_dist}
\end{figure}

\begin{figure}[h!]
    \centering
    \includegraphics[width=1\linewidth]{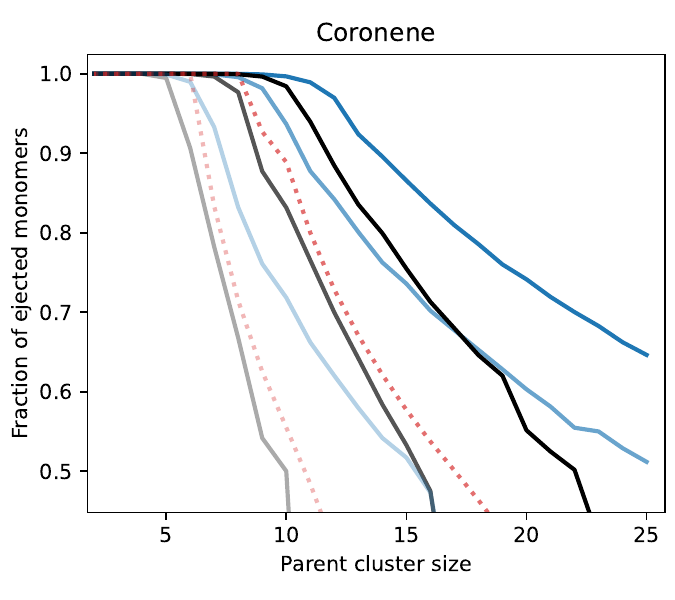}
    \caption{Fraction of evaporated cluster members after desorption from the cluster, similar to Figure \ref{fig:gasphase_cluster_evaporation}. The gas-phase cluster data is included as red dotted lines with the same colour scheme (light colour: 15\,eV, medium colour: 25\,eV, dark colour: 35\,eV excitation energy).}
    \label{fig:eject_members_desorbed}
\end{figure}

As described in Section \ref{sec:results_desorb}, modelled PAH clusters that desorb from the grain surface will have cooled down partially before entering the gas phase.
Figure \ref{fig:E_dist} shows the leftover energy distribution after desorption from the grain surface for a decamer PAH cluster adsorbed on a 400\,K dust grain and excited with 35\,eV.
The resulting gas-phase cluster can have energies below or above the assumed excitation energy of 35\,eV even though it has additional thermal energy compared to the case of gas-phase clusters described in Section \ref{sec:results_diss}.
Therefore, we can compare the number of ejected monomers from the cluster after desorption to the pure gas-phase calculations.
Figure \ref{fig:eject_members_desorbed} shows a comparison between gas-phase clusters and the cluster after desorption.

\section{Full cascade calculation}
In our model for coronene described in Section \ref{sec:discussion}, we estimate how a population of adsorbed PAH clusters will enter the gas-phase and will be dissociated through multiple X-rays.
The cascade can be described in multiple stages.
For clusters with fewer than ten monomers, the pathway to a monomer is described by
\begin{align}
\ce{(PAH)_p(s)} & \ce{->[X-ray] \text{$p$} PAH(g),}
\end{align}
where a cluster enters the gas-phase and entirely breaks down.
For clusters with between 11 and 21 members, the dissociation into monomers requires an additional X-ray after entering the gas-phase:
\begin{align}
\ce{(PAH)_m(s)} & \ce{->[X-ray] (PAH)_p(g) + (\text{$m$-$p$}) PAH(g)},\\
\ce{(PAH)_p(g)} & \ce{->[X-ray] \text{$p$} PAH(g).}
\end{align}
Finally, for the largest considered clusters, the dissociation cascade requires a third X-ray absorption:
\begin{align}
\ce{(PAH)_n(s)} & \ce{->[X-ray] (PAH)_m(g) + (\text{$n$-$m$}) PAH(g)},\\
\ce{(PAH)_m(g)} & \ce{->[X-ray] (PAH)_p(g) + (\text{$m$-$p$}) PAH(g)},\\
\ce{(PAH)_p(g)} & \ce{->[X-ray] \text{$p$} PAH(g).}
\end{align}
Table \ref{tab:coefficient_cascade} gives the desorption rate and corresponding intermediate cluster sizes $n,m,$ and $p$ for desorbed clusters.
The breakdown of a each species can be calculated using exponential decays of the form
\begin{equation}
    N(t) = N_0 \cdot \text{exp}(-kt)\cdot \text{.}
\end{equation}

\begin{table}[h!]
    \centering
    \caption{Cluster sizes and desorption rates after desorption from the grain surface. The larger the cluster, the more X-rays are required to fully dissociate the cluster into monomers.}
    \begin{tabular}{c c c c c c c c}
    \hline
    \multicolumn{4}{c}{Coronene} & \multicolumn{4}{c}{Ovalene}\\
    \hline
    $n$ & $m$ & $p$ & $k_\text{des} [1/\text{s}]$ & $n$ & $m$ & $p$ & $k_\text{des} [1/\text{s}]$\\ \hline\hline
     && 1 & $1.8 \cdot 10^{-9}$ & & & 1& $2.4 \cdot 10^{-9}$\\
     && 2 & $3.6 \cdot 10^{-9}$ & & & 2&$4.8 \cdot 10^{-9}$\\
     && 3 & $5.4 \cdot 10^{-9}$ & & & 3&$7.2 \cdot 10^{-9}$\\
     && 4 & $7.2 \cdot 10^{-9}$ & & & 4&$5.3 \cdot 10^{-9}$\\
     && 5 & $8.1 \cdot 10^{-9}$ & & & 5&$1.2 \cdot 10^{-9}$\\
     && 6 & $4.8 \cdot 10^{-9}$ & & & 6&$2.5 \cdot 10^{-10}$\\
     && 7 & $2.2 \cdot 10^{-9}$ & & & 7&$6.6 \cdot 10^{-11}$\\
     && 8 & $1.1 \cdot 10^{-9}$ & & & 8&$2.0 \cdot 10^{-11}$\\
     && 9 & $4.6 \cdot 10^{-10}$ & & & 9&$7.6 \cdot 10^{-12}$\\
     && 10 & $2.4 \cdot 10^{-10}$ & & 10 & 2&$2.9 \cdot 10^{-12}$\\
     & 11 & 2 &$1.5 \cdot 10^{-10}$ & & 11 & 3&$1.3 \cdot 10^{-12}$\\
     & 12 & 2 &$6.0 \cdot 10^{-11}$ & & 12& 4&$7.0 \cdot 10^{-13}$\\
     & 13 & 3 &$5.4 \cdot 10^{-11}$ & & 13& 5&$3.1 \cdot 10^{-13}$\\
     & 14 & 4 &$3.4 \cdot 10^{-11}$ & & 14& 7&$1.7 \cdot 10^{-13}$\\
     & 15 & 5 &$2.3 \cdot 10^{-11}$\\
     & 16 & 5 &$1.6 \cdot 10^{-11}$\\
     & 17 & 6 &$1.1 \cdot 10^{-11}$\\
     & 18 & 7 &$8.3 \cdot 10^{-12}$\\
     & 19 & 8 &$5.9 \cdot 10^{-12}$\\
     & 20 & 9 &$4.4 \cdot 10^{-12}$\\
     21 & 9 & 2 &$3.5 \cdot 10^{-12}$\\
     22 & 10 & 2 &$2.6 \cdot 10^{-12}$\\
     23 & 11 & 3 &$2.1 \cdot 10^{-12}$\\
     24 & 12 & 4 &$2.0 \cdot 10^{-12}$\\
     25 & 13 & 5 &$1.4 \cdot 10^{-12}$\\
     \hline
    \end{tabular}
    \label{tab:coefficient_cascade}
\end{table}
\end{appendix}

\end{document}